\documentclass[aps, prb, twocolumn, oneside, 10pt, amsfonts, amsmath, nofootinbib, floatfix, superscriptaddress]{revtex4-2}

\usepackage{graphicx}
\usepackage[colorlinks=true]{hyperref}
\usepackage{physics}

\usepackage{soul}

\usepackage{caption}
\usepackage{subcaption}
\usepackage{blkarray}
\usepackage{kbordermatrix} 
\renewcommand{\kbldelim}{(} 
\renewcommand{\kbrdelim}{)}

\usepackage{xcolor}
\usepackage{wasysym}
\DeclareMathSymbol{\mhexagon}{\mathord}{wasy}{55}

\newcommand{\fv}{{\bf f}}
\newcommand{\kv}{{\bf k}}

\newcommand{\rv}{{\bf r}}

\newcommand{\vv}{{\bf v}}

\newcommand{\n}[1]{{\bf n}^{(#1)}}

\newcommand{\av}[1]{{\bf a}^{(#1)}}
\newcommand{\bv}[1]{{\bf b}^{(#1)}}
\newcommand{\AV}[1]{{\bf A}^{(#1)}}
\newcommand{\BV}[1]{{\bf B}^{(#1)}}
\newcommand{\VV}{{\bf V}}

\newcommand{\CL}{{\cal L}}
\newcommand{\CT}{{\cal T}}
\newcommand{\CE}{{\cal E}}
\newcommand{\CR}{{\cal R}}
\newcommand{\Hzz}{H_{00}}
\newcommand{\Hhz}{H_{\frac{1}{2}0}}
\newcommand{\Hzh}{H_{0\frac{1}{2}}}
\newcommand{\Hhh}{H_{\frac{1}{2}\frac{1}{2}}}

\begin{document}
\title{Hexagonal and trigonal quasiperiodic tilings}
\author{Sam Coates}
\email{Corresponding author: samuel.coates@liverpool.ac.uk}
\affiliation{Department of Materials Science and Technology, Tokyo University of Science, Katsushika City, Tokyo 125--8585, Japan}
\affiliation{Surface Science Research Centre and Department of Physics, University of Liverpool, Liverpool L69 3BX, UK}
\author{Akihisa Koga}
\affiliation{Department of Physics, Tokyo Institute of Technology, Meguro, Tokyo 152--8551, Japan}
\author{Toranosuke Matsubara}
\affiliation{Department of Physics, Tokyo Institute of Technology, Meguro, Tokyo 152--8551, Japan}
\author{Ryuji Tamura}
\affiliation{Department of Materials Science and Technology, Tokyo University of Science, Katsushika City, Tokyo 125--8585, Japan}
\author{Hem Raj Sharma}
\author{Ronan McGrath}
\affiliation{Surface Science Research Centre and Department of Physics, University of Liverpool, Liverpool L69 3BX, UK}
\author{Ron Lifshitz}
\email{Corresponding author: ronlif@tau.ac.il}
\affiliation{Surface Science Research Centre and Department of Physics, University of Liverpool, Liverpool L69 3BX, UK}
\affiliation{Raymond and Beverly Sackler School of Physics and Astronomy, Tel Aviv University, Tel Aviv 69978, Israel}

\begin{abstract}
Exploring nonminimal-rank quasicrystals, which have symmetries that can be found in both periodic and aperiodic crystals, often provides new insight into the physical nature of aperiodic long-range order in models that are easier to treat. Motivated by the prevalence of experimental systems exhibiting aperiodic long-range order with hexagonal and trigonal symmetry, we introduce a generic two-parameter family of 2-dimensional quasiperiodic tilings with such symmetries. We focus on the special case of trigonal and hexagonal Fibonacci, or golden-mean, tilings, analogous to the well studied square Fibonacci tiling. We first generate the tilings using a generalized version of de Bruijn's dual grid method. We then discuss their interpretation in terms of projections of a hypercubic lattice from six dimensional superspace. We conclude by concentrating on two of the hexagonal members of the family, and examining a few of their properties more closely, while providing a set of substitution rules for their generation. 
\end{abstract}
\date{\today}
\maketitle

\section{Motivation}

Quasiperiodic tilings of the plane, like the famous decagonal Penrose tiling~\cite{Penrose74}, octagonal Ammann-Beenker tiling~\cite{Ammann92}, and dodecagonal Stampfli tiling~\cite{Stampfli86} have been used for decades to model 2-dimensional quasicrystals, or to model the high-symmetry surfaces of 3-dimensional quasicrystals. These tilings are attractive both mathematically and physically, partly because they possess symmetries that are incompatible with, or forbidden in, periodic crystals. Nevertheless, it has been argued~\cite{quasidef,*crystaldef} that no less important are those nonminimal-rank quasiperiodic tiling models, whose symmetries are not forbidden in periodic crystals~\cite{Sasisekharan1989non, Clark1991quasiperiodic}, such as the square Fibonacci tiling of rank 4, and its higher-dimensional generalizations~\cite{squarefib}. This was suggested even before such tilings had been reported experimentally, motivated by theoretical questions that are easier to solve on quasicrystalline tilings with such symmetries~\cite{Shahar06, *Shahar08, Damanik11}. Here we study the trigonal and hexagonal analogs of the square Fibonacci tiling, which possess a much richer structure.

To be clear with our terminology, recall that the density $\rho(\rv)$ of a \emph{quasiperiodic crystal} can be decomposed into countably many Fourier modes, 
\begin{equation}\label{Eq:crystal}
\rho(\rv) = \sum_{\kv \in \CL} \rho(\kv) e^{i\kv\cdot\rv},
\end{equation}
where the \emph{reciprocal lattice} $\CL$, consisting of the closure under addition of all wave vectors $\kv$ with nonzero Fourier coefficients $\rho(\kv)$, is a so-called finitely generated $\mathbb{Z}$-module. If the \emph{rank}, or smallest number $D$ of wave vectors required to generate $\CL$ over the integers, is equal to the spatial dimension $d$, then $\rho(\rv)$ is the density of a periodic crystal. If $D>d$, the reciprocal lattice $\CL$ is dense, and $\rho(\rv)$ is the density of a \emph{quasicrystal}, having in general no spatial translations that leave it invariant (see  Lifshitz~\cite{quasidef,*crystaldef} for further detail.). 

The rank of a crystal is required to have some minimum value, for the crystal to be compatible with a given point group symmetry. Pentagonal or decagonal crystals in two dimensions cannot be periodic, as they require a minimum rank of $D=4$. Trigonal or hexagonal crystals in two dimensions, on the other hand, require a minimum rank of $D=2$, and therefore may or may not be periodic. Here we are concerned with a family of nonminimal-rank---and therefore aperiodic---trigonal and hexagonal tilings with $D=4$. 

Nonminimal-rank quasicrystals do occur naturally, particularly on the low-symmetry 2-di\-men\-sion\-al surfaces of icosahedral quasicrystals, further motivating the development of appropriate tiling models. This was demonstrated very early on by \citet{Duneau85}, who generated 3-fold and 2-fold quasiperiodic tilings as 2-dimensional sections through a 3-dimensional icosahedral tiling model. In fact, a few of us~\cite{Coates18, Burnie20, Coates20} have identified the “hypothetical” square Fibonacci tiling~\cite{squarefib}, as well as a possible trigonal version of the tiling, while studying atomic and molecular overlayers on the 2-fold and 3-fold surfaces of icosahedral quasicrystals. Hexagonal quasicrystals appear also in incommensurate multilayer graphene~\cite{Woods14,Uri23,Oka21}, when the relative angles between the layers do not give rise to a periodic moir\'e pattern, nor are they equal to the 30$^\circ$ angle, required for increasing the symmetry from 6-fold to 12-fold. The resulting pattern is that of a 2-dimensional rank-4, or higher, hexagonal quasicrystal. Moreover, recent theoretical studies indicate that rank-4 hexagonal quasicrystalline patterns could also be stabilized on the surfaces of vibrated fluids~\cite{Iooss22}.

Motivated by these experimental systems exhibiting aperiodic long-range order with trigonal and hexagonal symmetry, we explore a two-parameter family of quasiperiodic trigonal and hexagonal tilings. We believe that the addition of such a generic family of tilings to the recent collection of ad-hoc aperiodic hexagonal tilings~\cite{Socolar11, Dotera17, Archer22} will facilitate more quantitative analysis of the 3-fold surfaces of icosahedral quasicrystals, and of the other systems mentioned above. We also note that two of us have already used one of the hexagonal members of this family of tilings to study the Hubbard model on a nonminimal-rank quasicrystal, and discovered ferrimagnetically ordered states~\cite{Koga22}.

We begin in Sec.~\ref{sec:context} by setting the broader context for the tilings studied here. In Sec.~\ref{sec:dual} we generate trigonal and hexagonal Fibonacci, or golden-mean, tilings using a generalized version of de Bruijn's dual grid method~\cite{deBruijn81, deBruijn86, Socolar85, Gahler86, Ho86, Rabson88, Rabson89, Lifshitz05}. We then discuss in Sec.~\ref{sec:projection} their interpretation in terms of projections of a hypercubic lattice from six dimensions.\footnote{For a pedagogical introduction to both of these methods see \citet{Senechal96}.} Finally, we focus in Sec.~\ref{sec:hex} on two of the hexagonal members of the family, describing a set of  substitution rules for their construction, while examining more closely a few of their properties such as their tile and vertex frequencies.

\section{Setting the context}
\label{sec:context}

We wish to create 2-dimensional quasiperiodic tilings whose diffraction patterns are supported on hexagonal reciprocal lattices $\CL$ of rank 4. It is always possible to generate such lattices by two mutually incommensurate 6-fold stars of wave vectors~\cite[Appendix]{Lifshitz94}. This requires that the two stars cannot both be generated from a single 6-fold star, thereby reducing the rank to 2. Each star, in turn, is generated by a pair of equal-length wave vectors, $\kv_1$ and $\kv_2$, separated by $120^\circ$, by taking $\pm\kv_1$, $\pm\kv_2$ and $\pm\kv_3 = \mp(\kv_1 + \kv_2)$. We shall denote the ratio of the magnitudes of two vectors, one from each star, by $\tau$, and denote the relative angle between the stars by $\theta$.

As described by \citet{Lifshitz94}, such lattices belong to three different Bravais classes, depending on the angle $\theta$ of their relative orientation.  If the vectors of one star are oriented along, or exactly between, the vectors of the other ($\theta=0^\circ$ or $30^\circ$), the point group is $6mm$, and the Bravais classes are denoted by $[4,0]$ and $[2,2]$ respectively, specifying the number of generating vectors along each type of mirror. Note that in the latter case, if the vectors of the two stars are of equal length ($\tau=1$), the symmetry increases to 12-fold, and the point group becomes $12mm$. For any other arbitrary angle $\theta$, the lattice loses its mirrors and the point group is reduced to 6. In this case the Bravais class is denoted by $[4]$, as there are no longer any mirrors with which to orient the stars. Here, if $\tau=1$, the point group remains $6mm$ with all six mirrors lying halfway between pairs of nearby generating vectors. This is the generic situation in bilayer graphene.

Finally, recall that the diffraction diagrams of trigonal crystals in 2-dimensions, whether periodic or not, appear to have hexagonal symmetry. This is because for any real-valued function, or density $\rho(\rv)$ as in Eq.~\eqref{Eq:crystal}, $\rho(-\kv)={\rho}^\ast(\kv)$, where the asterisk denotes complex conjugation. As a result, if $\rho(\rv)$ has 3-fold symmetry, the magnitudes of its Fourier coefficients $|\rho(\kv)|$, and therefore its diffraction pattern, will necessarily have 6-fold symmetry. The distinction between 2-dimensional trigonal and hexagonal crystals is thus encoded in the phases of the Fourier coefficients.

Tilings with any value of $\tau$ and $\theta$ can be generated using the dual grid method, or its interpretation in terms of projections from 6-dimensional space, as described in sections \ref{sec:dual} and \ref{sec:projection} below. We choose to concentrate here on the case, most relevant for the 3-fold surfaces of icosahedral crystals, where both stars are in the same orientation ($\theta=0^\circ$), and the length ratio is the golden mean $\tau=(1+\sqrt{5})/2$. Thus, in terms of the unit vectors
\begin{equation}\label{Eq:unitvecs}
    \n{j}=\left(\cos{\frac{2\pi(j-1)}{3}}, \sin\frac{2\pi(j-1)
    }{3}\right),\quad j\in {\mathbb Z},
\end{equation}
of which only three are unique, and following the notation of \citet{Rabson88,Rabson89}, the two 6-fold stars are given by the six wave vectors
\begin{equation}\label{Eq:k-star}
    \kv^{(j)}=\frac{2\pi}{L_j}\n{j},\quad j=1,\ldots 6,
\end{equation}
and their negatives, where
\begin{equation}\label{Eq:Lengths}
    L_j=\begin{cases}
    \tau, & j=1,2,3,\\
    1, & j=4,5,6.
    \end{cases}
\end{equation}
Many of the general properties of the tilings that we study here are valid for any irrational $\tau$, yet some special properties are true only for the metallic means, which satisfy the relation $\tau^2=n\tau+1$ ($n\in {\mathbb N}$), or equivalently $\tau^{-1}=\tau-n$, which we use throughout with $n=1$ for the golden mean. These are mentioned again in Sec.~\ref{sec:summary} below, and treated by substitution rules elsewhere~\cite{Matsubara23}.

\section{The dual grid method}
\label{sec:dual}

To generate the tilings we use a generalized version of de Bruijn's dual grid method~\cite{deBruijn81, deBruijn86, Socolar85, Gahler86, Ho86, Rabson88, Rabson89, Lifshitz05}. To do so we associate with each of the six wave vectors $\kv^{(j)}$ in Eq.~\eqref{Eq:k-star}---now called \emph{grid vectors}---a corresponding \emph{tiling vector} $\av{j}$, to be determined below, and a grid consisting of an infinite family of equally spaced parallel lines. The lines of the $j^{th}$ grid are normal to the unit vector $\n{j}$, separated by a distance $L_j$, and shifted from the origin (in the direction of $-\n{j}$) by an amount $f_jL_j$, for some chosen set of \emph{grid shifts} $0\le f_j<1$.\footnote{The $j^{th}$ grid can be thought of as the peaks of the plane wave, or Fourier mode, given by the function $\cos{(\kv^{(j)}\cdot\rv+f_j)}$, where the grid shift acts as a phase shift of the Fourier mode.} These six grids together generate the \emph{dual grid} or \emph{multigrid}---in this case a superposition of a pair of trigrids, or a \emph{double trigrid} for short---which divides grid space into cells. 
Figure \ref{Fig:grids} shows several examples of single and double trigrids, generated by choosing different grid shifts $f_j$. 

The double trigrid is dual to the tiling in the sense that each {\it intersection\/} of lines in the grid corresponds to a {\it tile\/} in the tiling, whose edges are the tiling vectors $\av{j}$ associated with the families of the intersecting lines. In addition, each {\it cell\/} in the dual grid, labeled by six integers $n_j$, determines a {\it vertex\/} in the tiling at position $\sum_{j} n_j \av{j}$, where $n_j$ is the (signed) number of lines of the $j^{th}$ family, separating the cell from the origin (in the positive or negative direction of $\n{j}$). Only those integer linear combinations that correspond to cells in the dual grid are included as vertices in the tiling. Thus, it is the topology of the dual grid---the order in which grid lines intersect---that encodes the information needed for constructing the tiling.

A dual grid is said to be \emph{regular} if no more than two lines intersect at any point, and is otherwise considered \emph{singular}. An edge, formed by a tiling vector $\av{j}$, is included in the tiling whenever a line from the $j^{th}$ grid is crossed in going from one cell to another. Thus, regular grids produce tilings containing only parallelogram-shaped tiles that may become rhombs, rectangles, or squares. On the other hand, the tilings produced by singular grids contain tiles with more than 4 edges. 

\begin{figure}
    \centering
    \begin{subfigure}[b]{0.465\linewidth}
         \centering
         \includegraphics[width=\linewidth]{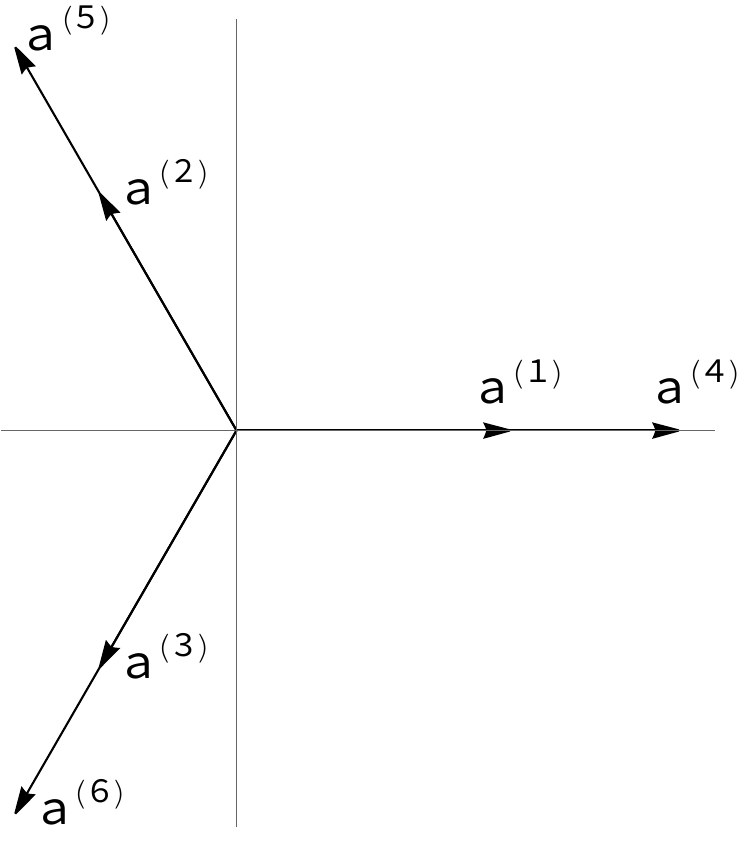}
         \caption{Tiling-space vectors}
         \label{fig:avecs}
     \end{subfigure}
     \begin{subfigure}[b]{0.452\linewidth}
         \centering
         \includegraphics[width=\linewidth]{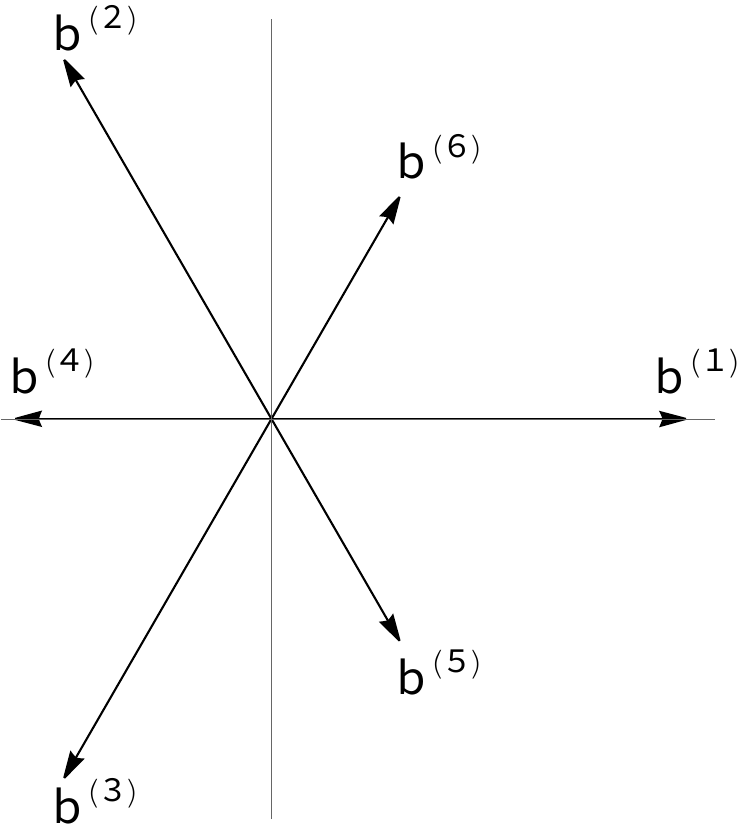}
         \caption{Internal-space vectors}
         \label{fig:bvecs}
     \end{subfigure}
	\caption{(a)~The tiling vectors of Eq.~\eqref{Eq:a-star}, where the vectors $\av{1}$, $\av{2}$, and $\av{3}$, are a factor of $\tau$ \emph{shorter} than the the vectors $\av{4}$, $\av{5}$, and $\av{6}$. The grid vectors $\kv^{(j)}$ given by Eq.~\eqref{Eq:k-star} are proportional to the tiling vectors $\av{j}$. (b)~The internal-space vectors given by Eq.~\eqref{Eq:b-star}, where the vectors $\bv{1}$, $\bv{2}$, and $\bv{3}$, are a factor of $\tau$ \emph{longer} than the the vectors $\bv{4}$, $\bv{5}$, and $\bv{6}$, and point in the opposite direction.
	\label{Fig:vectors}}
\end{figure}

A canonical choice of the six 2-dimensional tiling vectors, $\av{j}$, which is convenient for the analysis that is to follow later, and particularly for the calculation of the Fourier spectrum~\cite[section IV.C]{Rabson88}, is one that satisfies the so-called Ho condition~\cite{Ho86},
\begin{equation}\label{Eq:Ho}
    \sum_{j=1}^6 \av{j}_\mu \kv^{(j)}_\nu = 2\pi \delta_{\mu\nu},\quad \mu,\nu=1,2.
\end{equation}
We choose to satisfy the Ho condition by taking
\begin{equation}\label{Eq:a-star}
    \av{j}
    =\frac{2}{3(1+\tau^{-2})}\frac{1}{L_j} \n{j}
    =\frac{2\tau}{3\sqrt{5}}\frac{1}{L_j} \n{j},\quad j=1,\ldots 6,
\end{equation}
where the last equality holds if $\tau$ is the golden mean, or for any of the other metallic means by replacing $\sqrt{5}$ with $\sqrt{n^2+4}$. This choice takes the tiling vectors, shown in Fig.~\ref{fig:avecs}, to be proportional to the grid vectors, and inversely proportional to the separations $L_j$ between grid lines. This has the effect that the grid lines associated with long tiling vectors, $\av{4}, \av{5}$, and $\av{6}$, are crossed more frequently than the ones associated with short tiling vectors, $\av{1}, \av{2}$, and $\av{3}$. Consequently, large tiles are expected to appear more frequently in the tiling than small tiles. This is the convention for Fibonacci tilings (compare with the square Fibonacci case~\cite{squarefib}). 

All that is left is to understand what happens when one varies the grid shifts $f_j$. Because we are dealing with rank-4 tilings, there are four linearly independent combinations of grid shifts that do not alter the topology of the grid and leave the tiling indistinguishable.\footnote{According to~\citet{RWM88a,RWM88b}, two crystals are \emph{indistinguishable} if their densities share the same $n$-point autocorrelation functions, for any $n$. If two periodic crystals are indistinguishable they differ by no more than a translation. If the crystals are aperiodic, the relative phases of their Fourier modes $\rho(\kv)$ may differ more generally, affecting a so-called \emph{phason displacement}. For pedagogical overviews of the notion of indistinguishability, or local-isomorphism as it is called in the context of tiling, see \citet{Mermin92} or \citet{symbreak}.} These can be seen by considering the two trigonal subgrids as two rigid objects that can be shifted around in grid space. Displacing the two subgrids together along the $x$- or $y$-axes, relative to the origin of grid space, induces a rigid translation of the tiling. Displacing them in opposite directions along the $x$- or $y$-axes in grid space induces a so-called phason displacement of the tiling. 

This leaves two independent linear combinations of the six grid shifts that do affect the grid, and can potentially alter its topology and therefore change the nature of the tiling. These are the two sums of grid shifts within each of the individual subgrids, 
\begin{equation}\label{Eq:invariants}
    \alpha_s=f_1+f_2+f_3\quad \textrm{and}\quad \alpha_l=f_4+f_5+f_6,
\end{equation}
associated with independent combinations of grid vectors \eqref{Eq:k-star} that add up to zero. The corresponding sums of Fourier-mode phases are known in crystallography as \emph{structure invariants}. It can be shown, using the arguments of~\citet{Rabson89}, that if $\alpha_s$ and $\alpha_l$ are restricted to 0 or $1/2$ (modulo the integers), the dual grid and therefore also the resulting tiling are expected to have hexagonal symmetry. If at least one of the structure invariants is not equal to 0 or $1/2$  (modulo the integers), the tiling is expected to be trigonal only. 

\begin{figure}
    \centering
    \begin{subfigure}[b]{0.32\linewidth}
         \centering
         \includegraphics[width=\linewidth]{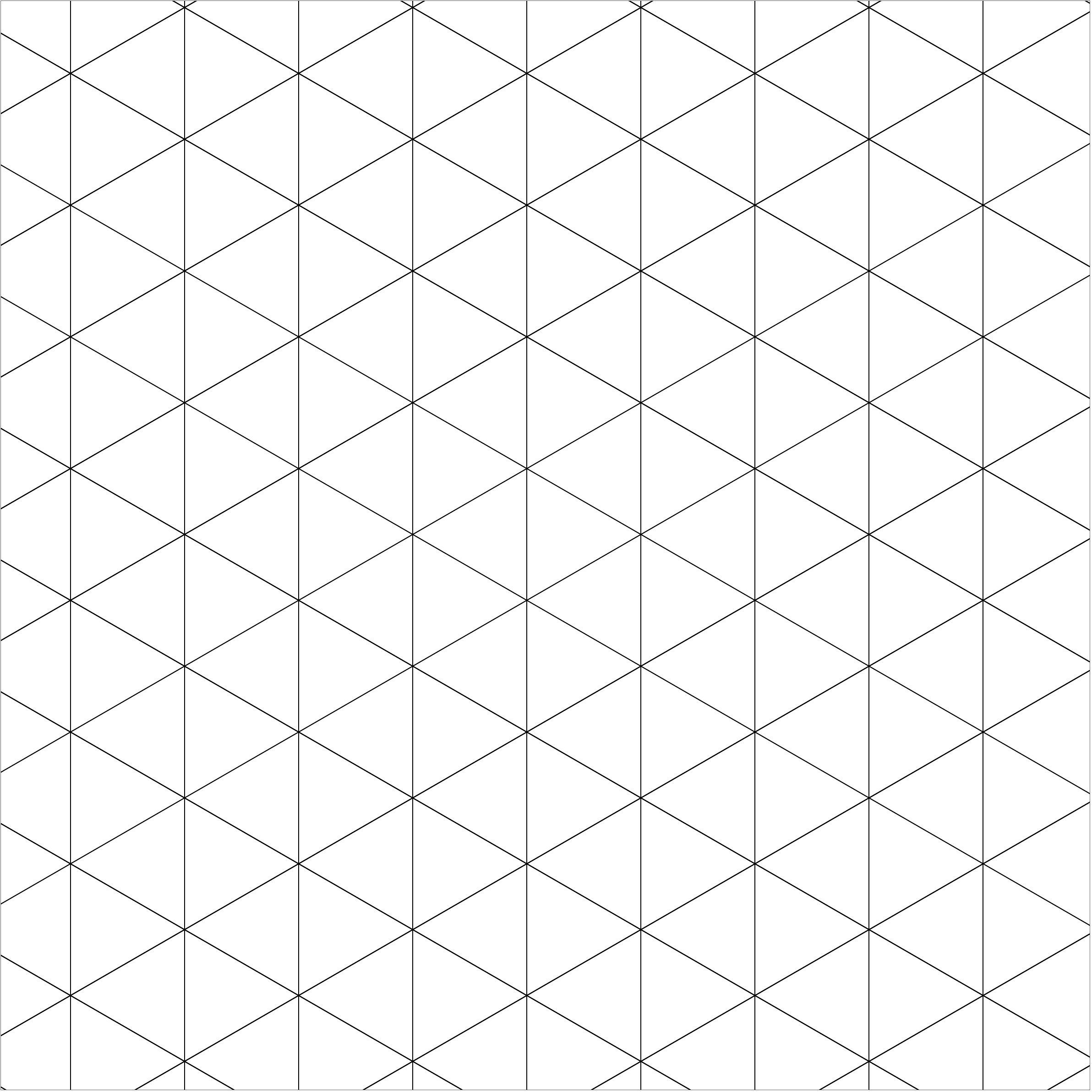}
         \caption{$\alpha\equiv0$}
         \label{fig:tri00}
     \end{subfigure}
     \begin{subfigure}[b]{0.32\linewidth}
         \centering
         \includegraphics[width=\linewidth]{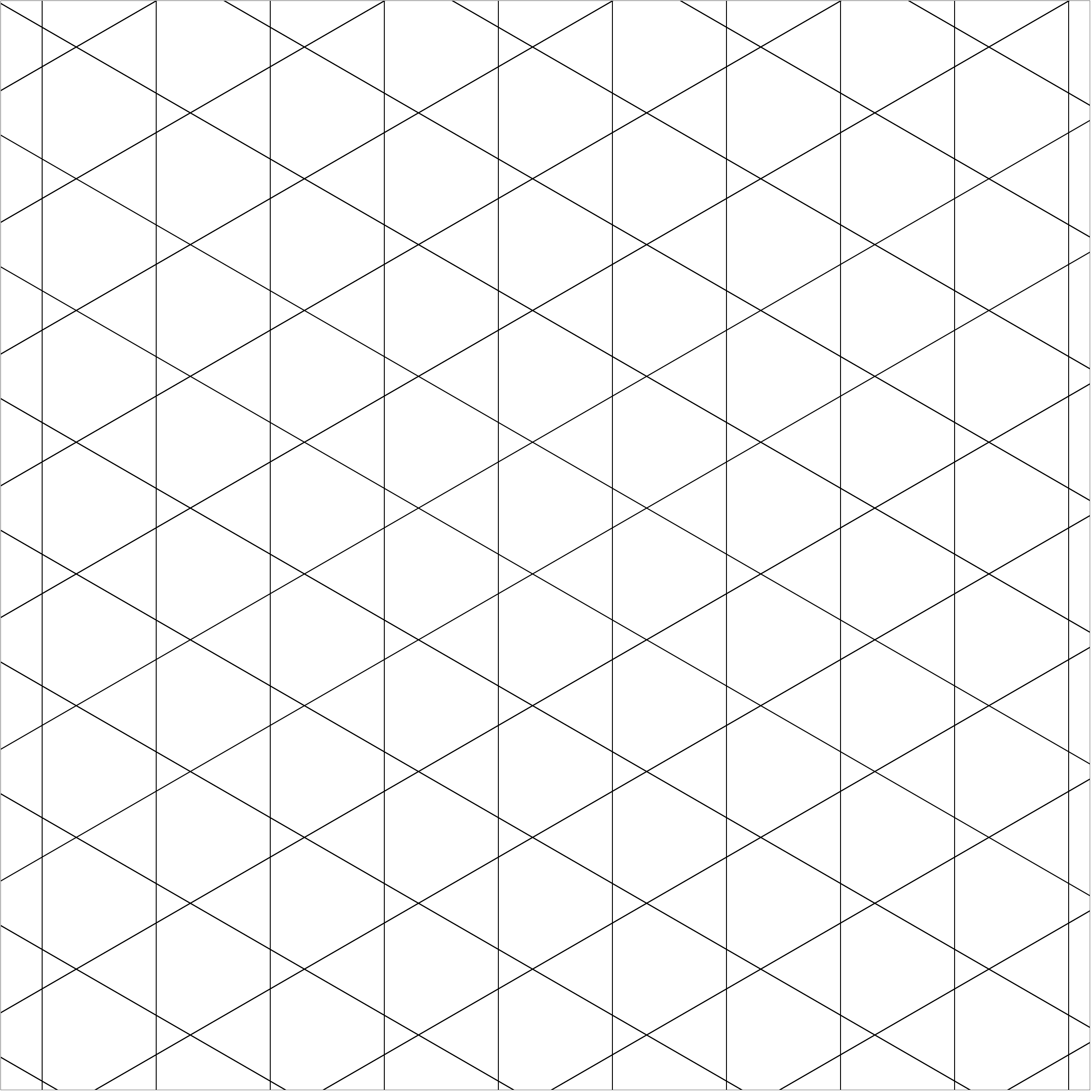}
         \caption{$\alpha\equiv.3$}
         \label{fig:tri03}
     \end{subfigure}
     \begin{subfigure}[b]{0.32\linewidth}
         \centering
         \includegraphics[width=\linewidth]{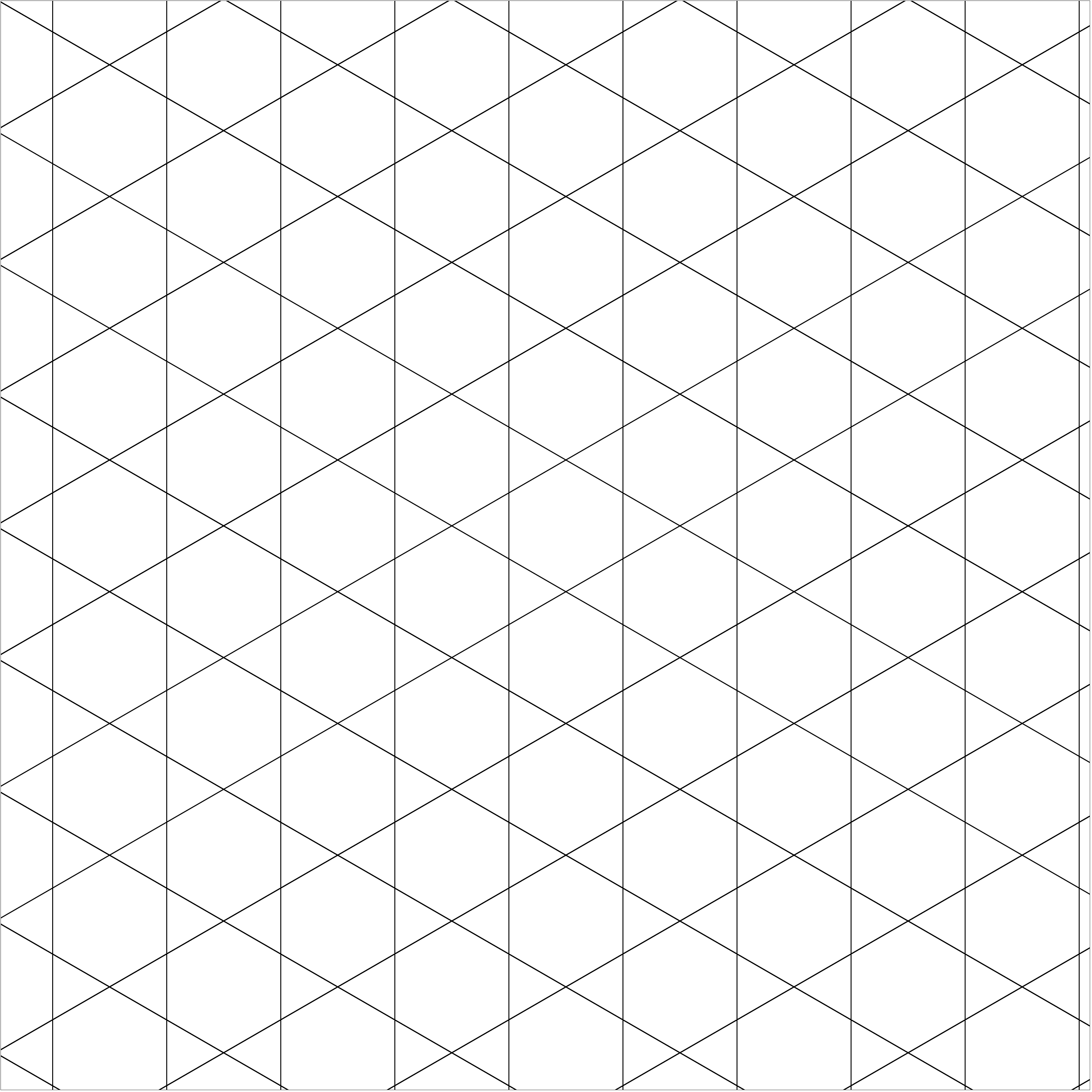}
         \caption{$\alpha\equiv.5$}
         \label{fig:tri05}
     \end{subfigure}
     \begin{subfigure}[b]{0.32\linewidth}
         \centering
         \includegraphics[width=\linewidth]{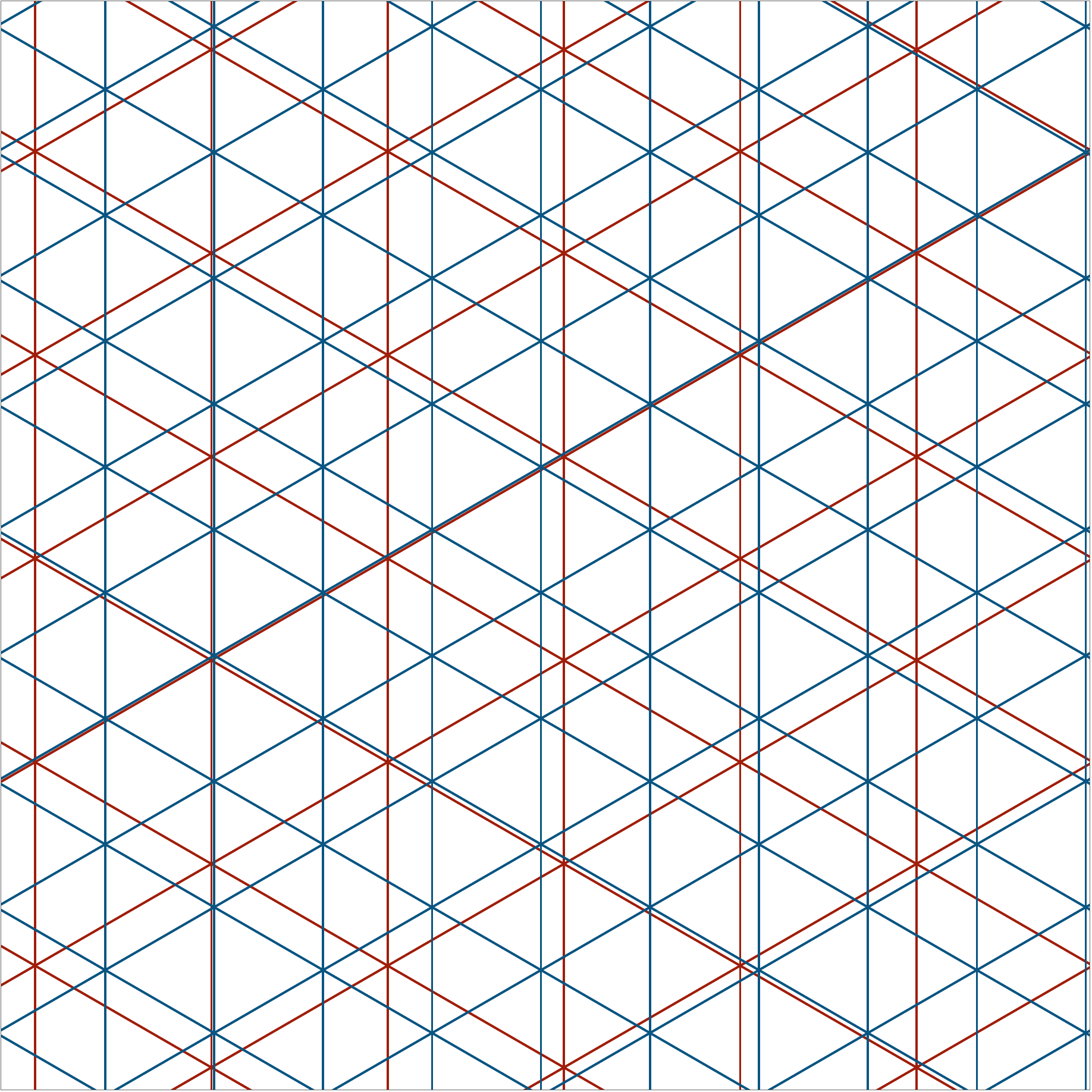}
         \caption{$\alpha_s\equiv\alpha_l\equiv0$}
         \label{fig:d-tri0000}
     \end{subfigure}
     \begin{subfigure}[b]{0.32\linewidth}
         \centering
         \includegraphics[width=\linewidth]{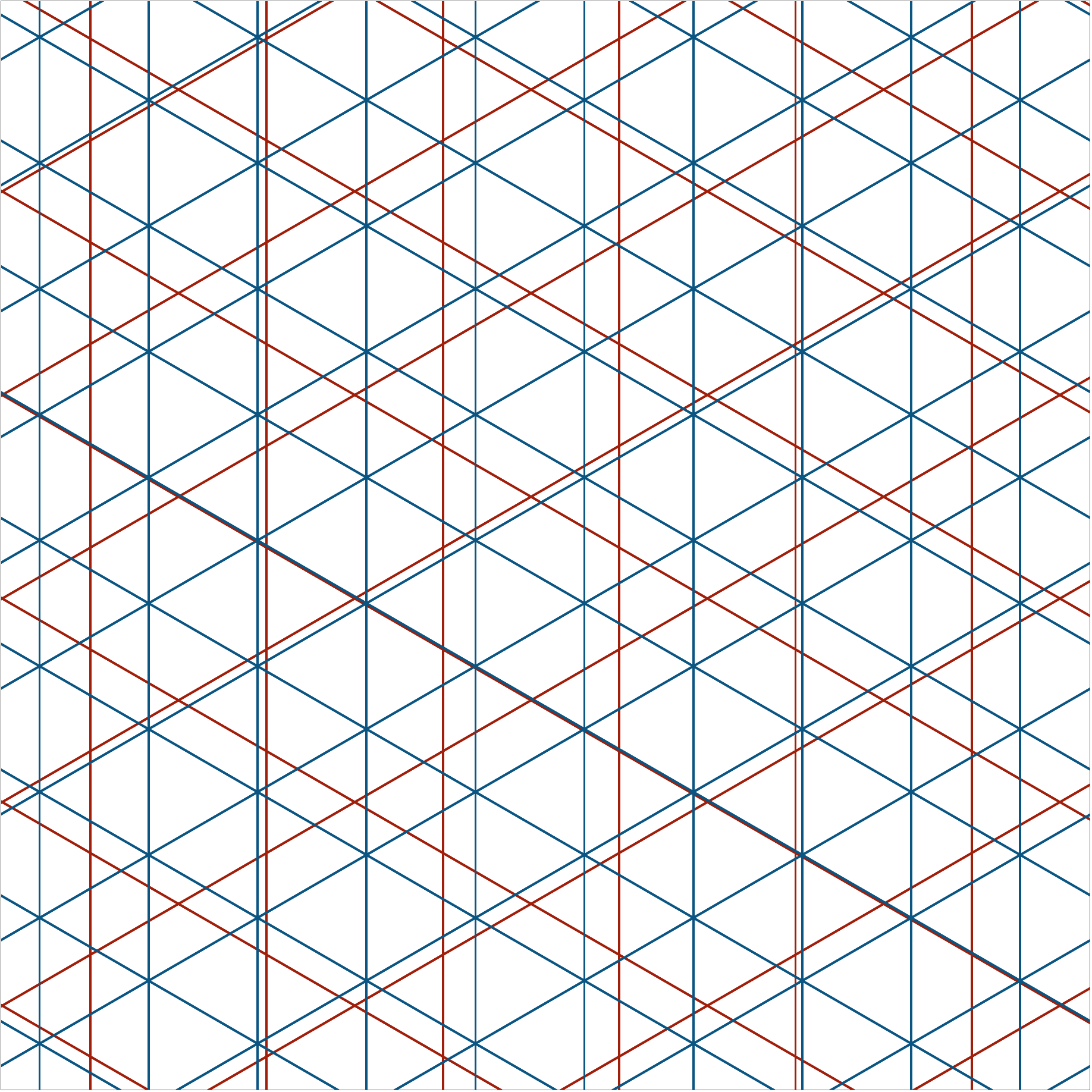}
         \caption{$\alpha_s\equiv.5$, $\alpha_l\equiv0$}
         \label{fig:d-tri0500}
     \end{subfigure}
     \begin{subfigure}[b]{0.32\linewidth}
         \centering
         \includegraphics[width=\linewidth]{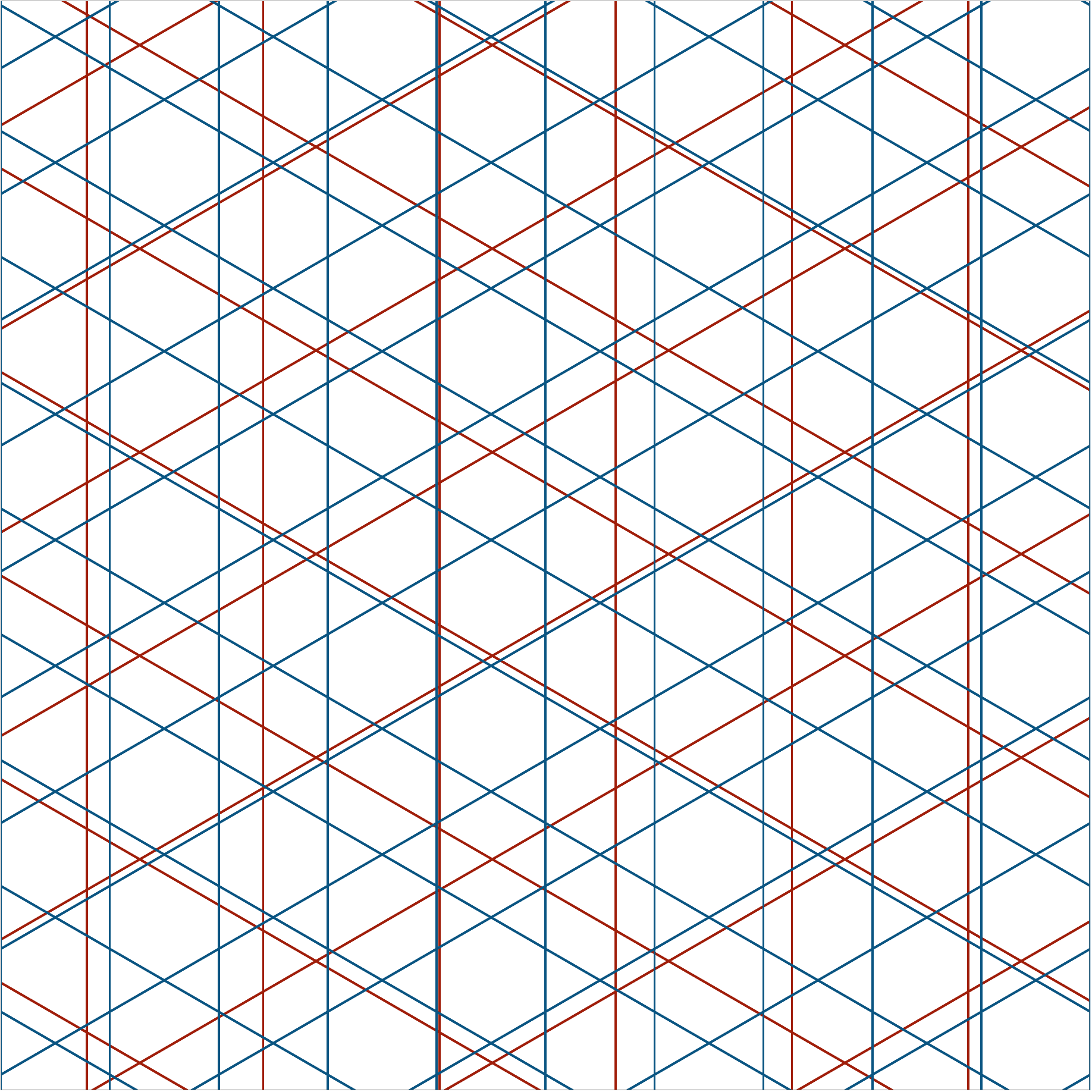}
         \caption{$\alpha_s\equiv.3$, $\alpha_l\equiv.5$}
         \label{fig:d-tri0305}
     \end{subfigure}
    \caption{Single trigrids (top row) and double trigrids (bottom row), for different values of the grid shifts $f_j$, as given by the structure invariants of Eq.~\eqref{Eq:invariants}, where ``$\equiv$'' denotes equality modulo the integers. Note that the grid is regular only if both structure invariants are nonzero. If one or both are zero the grid is singular or doubly singular, with triplets of lines intersecting on one or two triangular lattices of points, respectively.}
    \label{Fig:grids}
\end{figure}

The top row of Fig.~\ref{Fig:grids} shows single trigrids with different grid shifts. When their sum $\alpha\equiv0$, where ``$\equiv$'' denotes equality modulo the integers, the grid takes the form of a triangular lattice, with three grid lines intersecting at every lattice point. This singular grid is dual to a honeycomb tiling, consisting of hexagonal tiles only. When $\alpha\equiv0.5$, the grid takes the form of a kagome tiling, composed of regular hexagons and triangles, with no more than two lines intersecting at a point. For all other values of $\alpha$, the grid consists of irregular hexagons and two types of triangles, but its topology is equivalent to that of the kagome tiling. Thus, in the periodic case, all regular trigrids, for any $\alpha\not\equiv0$, produce the same periodic tiling---a rhombic tiling with hexagonal symmetry.

The bottom row of Fig.~\ref{Fig:grids} shows double trigrids with different grid shifts. If the structure invariant associated with a given trigonal subgrid is zero, its associated grid lines will form a triangular lattice, as in the single trigrid of Fig.~\ref{fig:tri00}, with triplets of intersecting lines. The resulting tiling is expected to contain hexagonal tiles. When both structure invariants are nonzero, lines within each trigonal subgrid intersect in pairs only, producing rhombic tiles. One can then shift the subgrids with respect to each other to ensure that their mutual intersections contain only pairs of lines. This will further introduce parallelogram tiles, with a pair of long edges and a pair of short edges.

\begin{figure*}[p]
    \centering
    \begin{subfigure}[b]{0.48\textwidth}
         \centering
         \includegraphics[width=\textwidth]{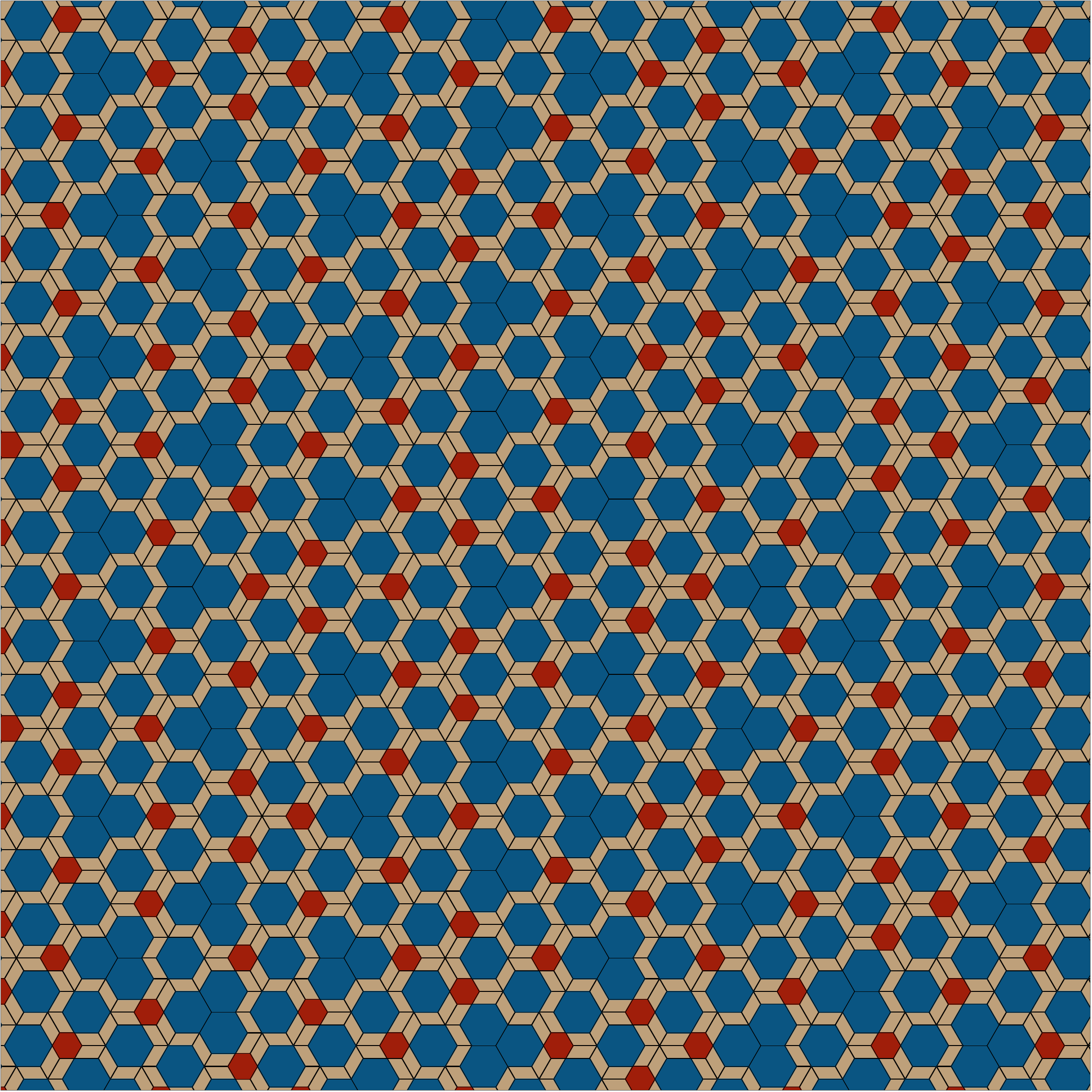}
         \caption{Hexagonal $\Hzz$ tiling with $\alpha_s\equiv\alpha_l\equiv0$}
         \label{fig:Hexa00}
     \end{subfigure}
     \hfill
     \begin{subfigure}[b]{0.48\textwidth}
         \centering
         \includegraphics[width=\textwidth]{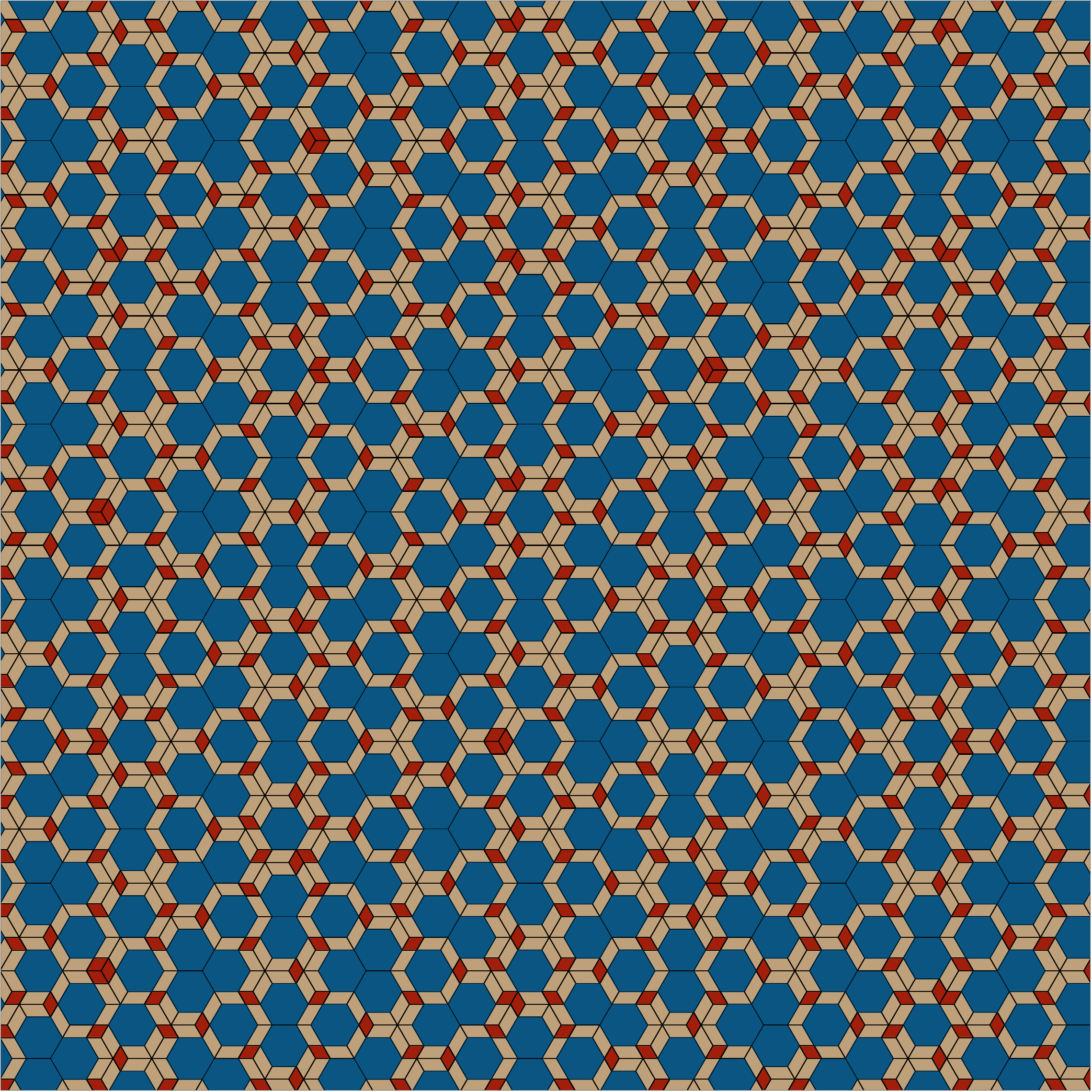}
         \caption{Hexagonal $\Hhz$ tiling with $\alpha_s\equiv0.5$, $\alpha_l\equiv0$}
         \label{fig:Hexa50}
     \end{subfigure}
     \vskip5pt
     \begin{subfigure}[b]{0.48\textwidth}
         \centering
         \includegraphics[width=\textwidth]{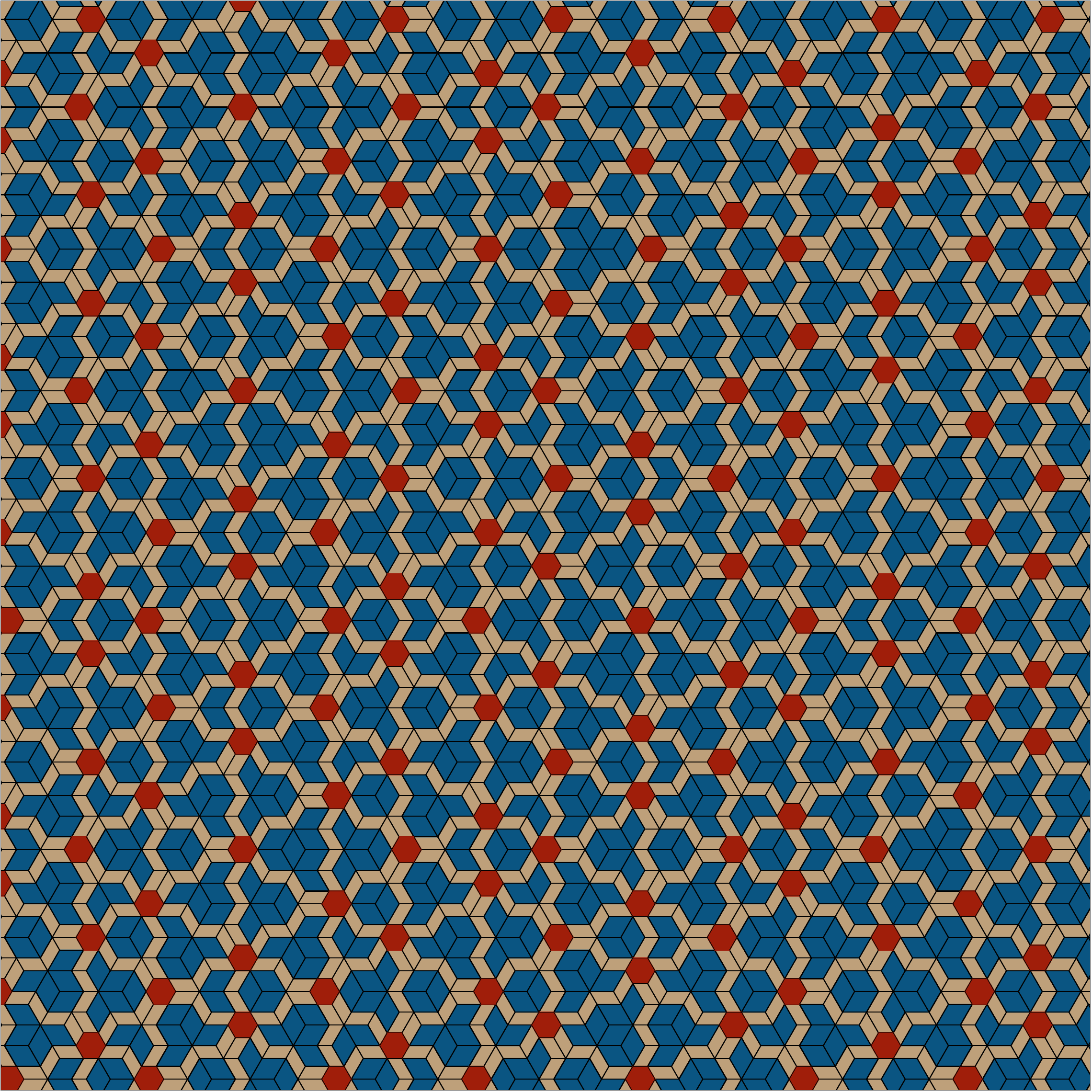}
         \caption{Hexagonal $\Hzh$ tiling with $\alpha_s\equiv0$, $\alpha_l\equiv0.5$}
         \label{fig:Hexa05}
     \end{subfigure}
     \hfill
     \begin{subfigure}[b]{0.48\textwidth}
         \centering
         \includegraphics[width=\textwidth]{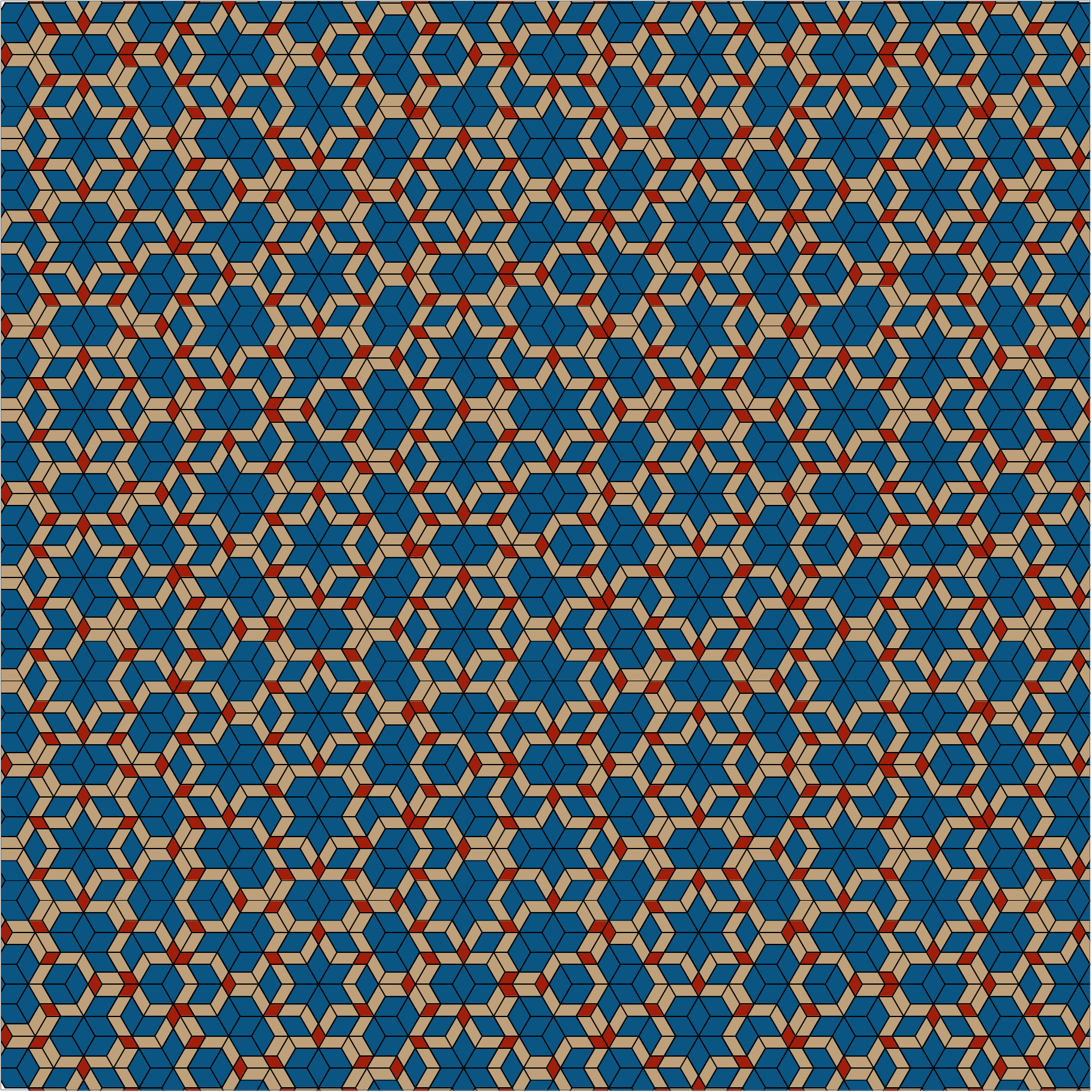}
         \caption{Hexagonal $\Hhh$ tiling with $\alpha_s\equiv\alpha_l\equiv0.5$}
         \label{fig:Hexa55}
     \end{subfigure}
    \caption{Hexagonal Fibonacci tilings, obtained by the dual grid method. (a)~With both $\alpha_s$ and $\alpha_l$ equal to zero (modulo the integers) the tiling is doubly singular, consisting of small and large hexagonal tiles and parallelograms. (b)~With only $\alpha_l\equiv0$, the tiling is still singular consisting of large hexagons, small rhombs, and parallelograms. (c)~With only $\alpha_s\equiv0$, the tiling remains singular, consisting of small hexagons, large rhombs, and parallelograms. (d)~With $\alpha_s\equiv\alpha_l\equiv0.5$ the tiling is regular, consisting of small and large rhombic tiles and parallelograms.}
    \label{Fig:Hexa-tilings}
\end{figure*}

\begin{figure*}
    \centering
    \begin{subfigure}[b]{0.48\textwidth}
         \centering
         \includegraphics[width=\textwidth]{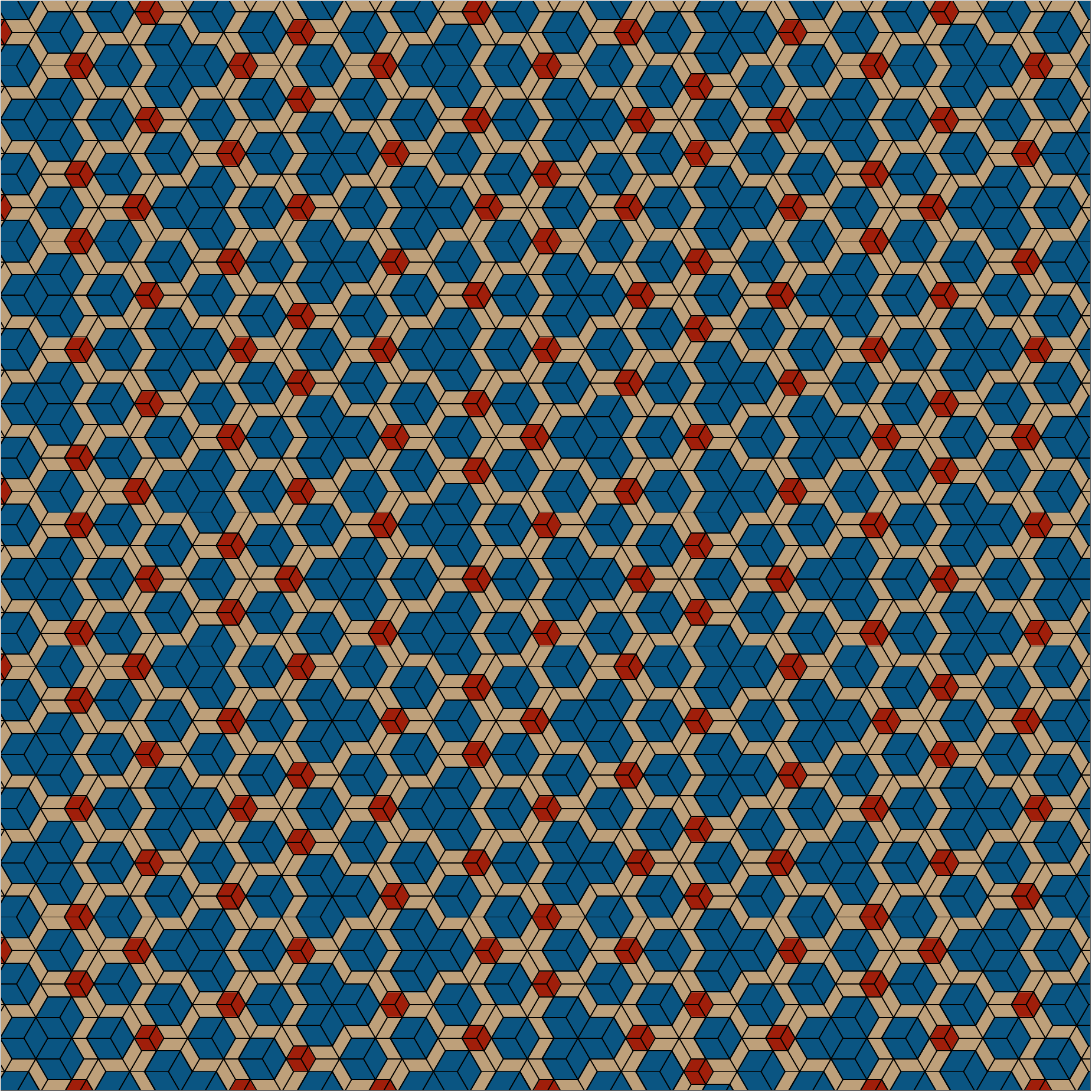}
         \caption{Trigonal tiling with $\alpha_s\equiv\alpha_l\equiv 10^{-8}$}
         \label{fig:Trig00}
     \end{subfigure}
     \hfill
     \begin{subfigure}[b]{0.48\textwidth}
         \centering
         \includegraphics[width=\textwidth]{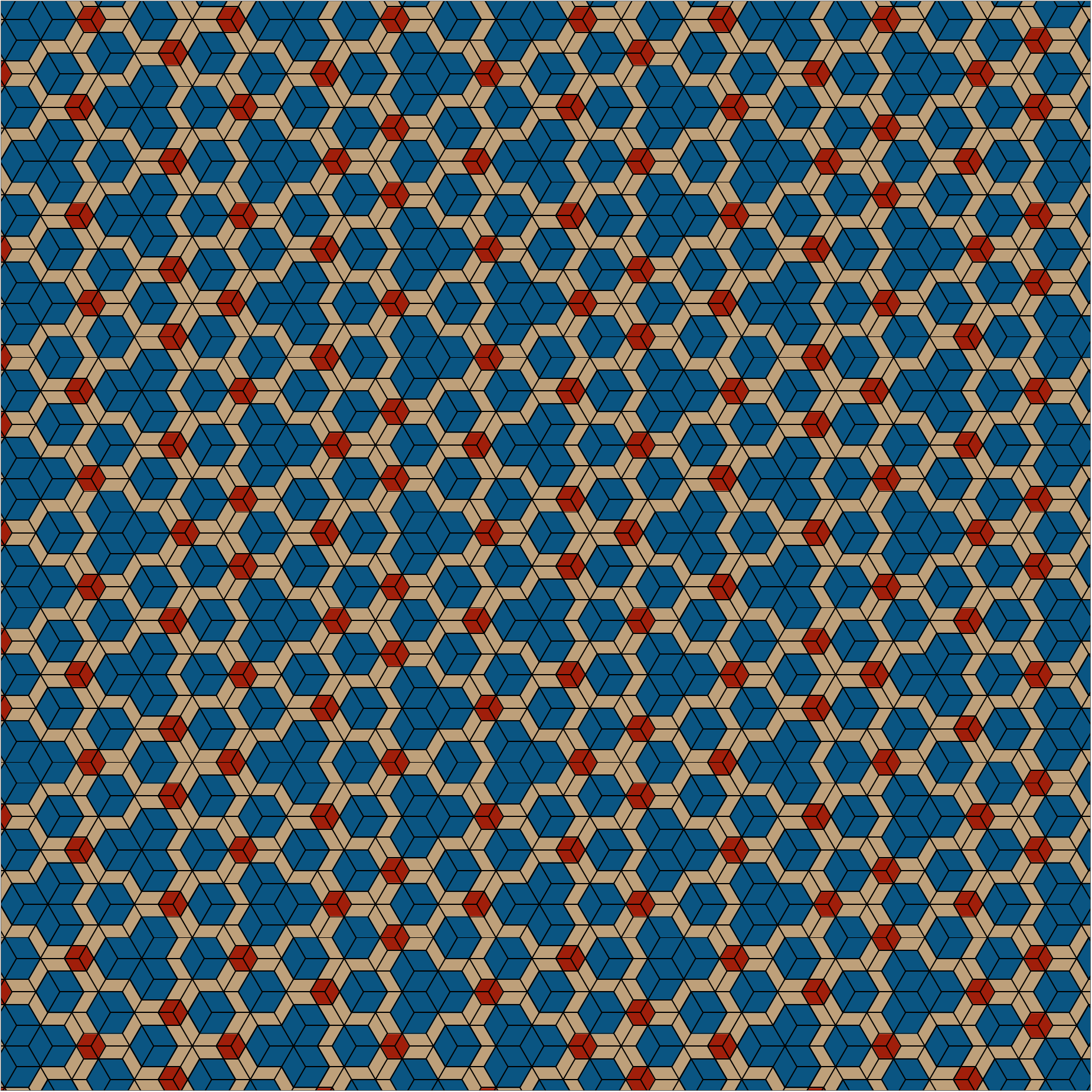}
         \caption{Trigonal tiling with $\alpha_s\equiv 10^{-8}$, $\alpha_l\equiv -10^{-8}$}
         \label{fig:Trig0-0}
     \end{subfigure}
     \vskip5pt
     \begin{subfigure}[b]{0.48\textwidth}
         \centering
         \includegraphics[width=\textwidth]{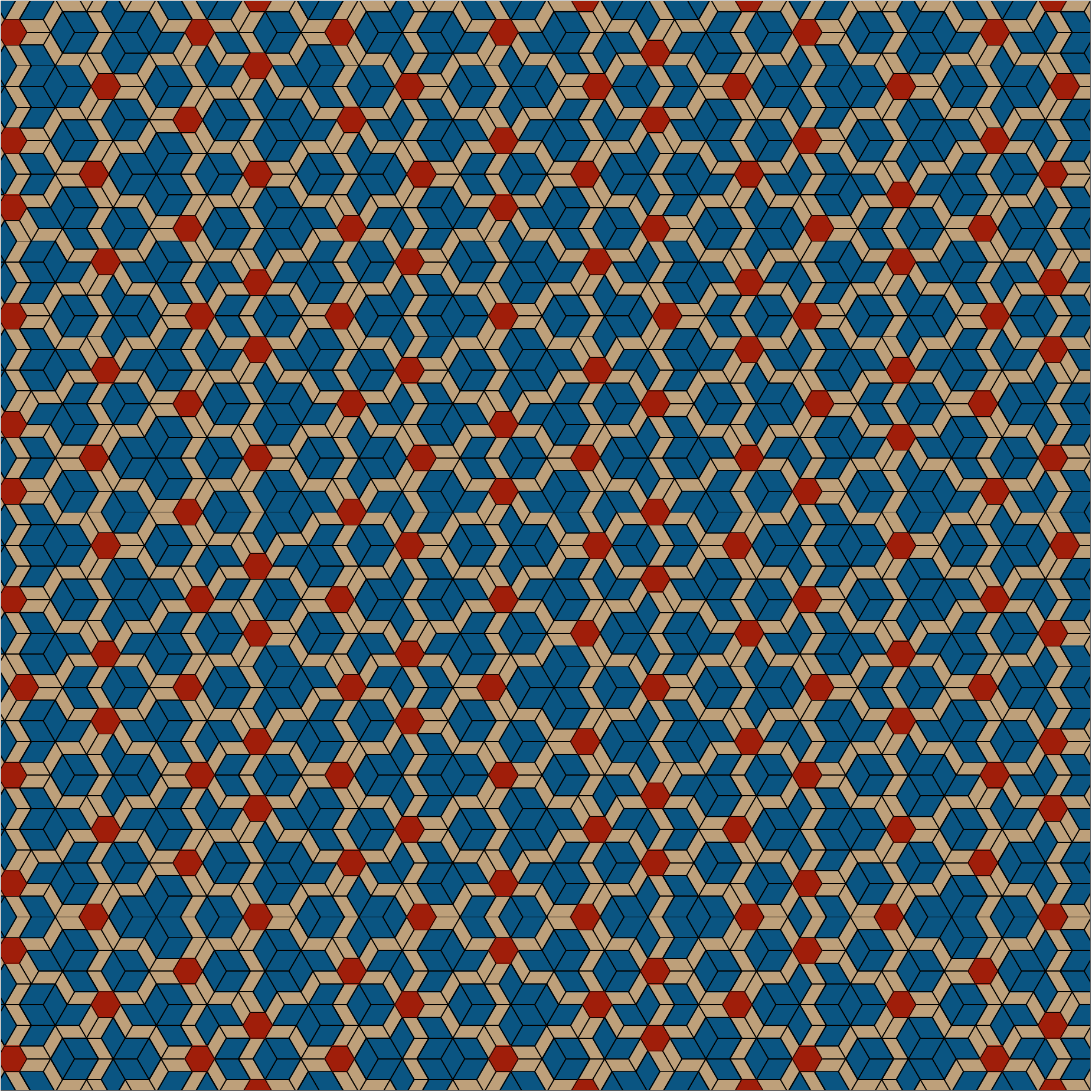}
         \caption{Trigonal tiling with $\alpha_s\equiv0$, $\alpha_l\equiv0.7$}
         \label{fig:Trig07}
     \end{subfigure}
     \hfill
     \begin{subfigure}[b]{0.48\textwidth}
         \centering
         \includegraphics[width=\textwidth]{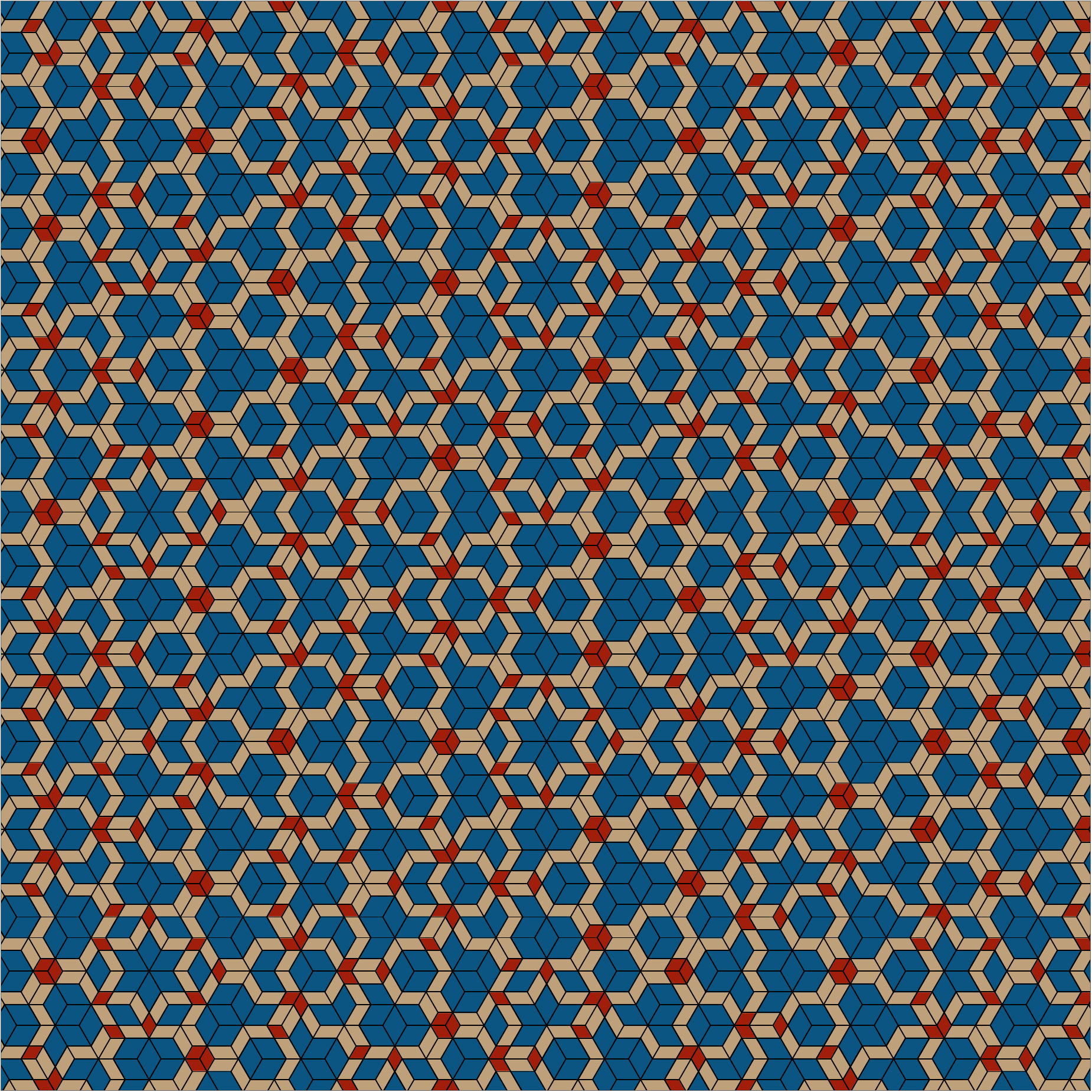}
         \caption{Trigonal tiling with $\alpha_s\equiv0.2$, $\alpha_l\equiv0.7$}
         \label{fig:Trig27}
     \end{subfigure}
    \caption{Trigonal Fibonacci tilings, obtained by the dual grid method. (a) With both $\alpha_s$ and $\alpha_l$ nearly equal to zero (modulo the integers) the tiling is almost identical to the hexagonal $\Hzz$ Fibonacci tiling of Fig.~\ref{fig:Hexa00}, except for the fact that all the hexagonal tiles are subdivided into rhombs that break the 6-fold symmetry and render the tiling trigonal. (b) With both $\alpha_s$ and $\alpha_l$ nearly equal to zero (modulo the integers), but with opposite sign, the large and small hexagons are subdivided into rhombs in opposite orientations. (c) With $\alpha_s\equiv0$, the tiling is singular, consisting of small hexagons, arranged as in the hexagonal tiling of Fig.~\ref{fig:Hexa05}. It is the arrangement of the large rhombs that breaks the 6-fold symmetry. (d) With both $\alpha_s$ and $\alpha_l$ far from 0 and $0.5$, we find individual small rhombs and pairs of small rhombs, in addition to triplets forming hexagons. Compare also the increased variety of patches of large rhombs with the tiling of Fig.~\ref{fig:Trig07}.}
    \label{Fig:Trig-tilings}
\end{figure*}

Figure~\ref{Fig:Hexa-tilings} confirms this analysis showing the four hexagonal Fibonacci tilings, labeled as $H_{\alpha_s,\alpha_l}$, obtained using the dual grid method by applying the four different assignments of 0 or $0.5$ to the two structure invariants $\alpha_s$ and $\alpha_l$. One can see that whenever a structure invariant is zero, the corresponding rhombic tile is replaced by a hexagonal one. These Fibonacci tilings all have the 2-dimensional hexagonal rank-4 symmorphic space group $[4,0]6mm$. Figure~\ref{Fig:Trig-tilings} shows a few examples of trigonal Fibonacci tilings with space group $[4,0]3m1$. One can see that for very small deviations of a structure invariant from 0, the corresponding hexagonal tiles are each subdivided into three rhombs in a manner that breaks the overall 6-fold symmetry, but otherwise nothing changes. We shall reinterpret this result later, in Sec.~\ref{sec:projection}, using the projection scheme.

\begin{figure*}
    \centering
    \includegraphics[width=\linewidth]{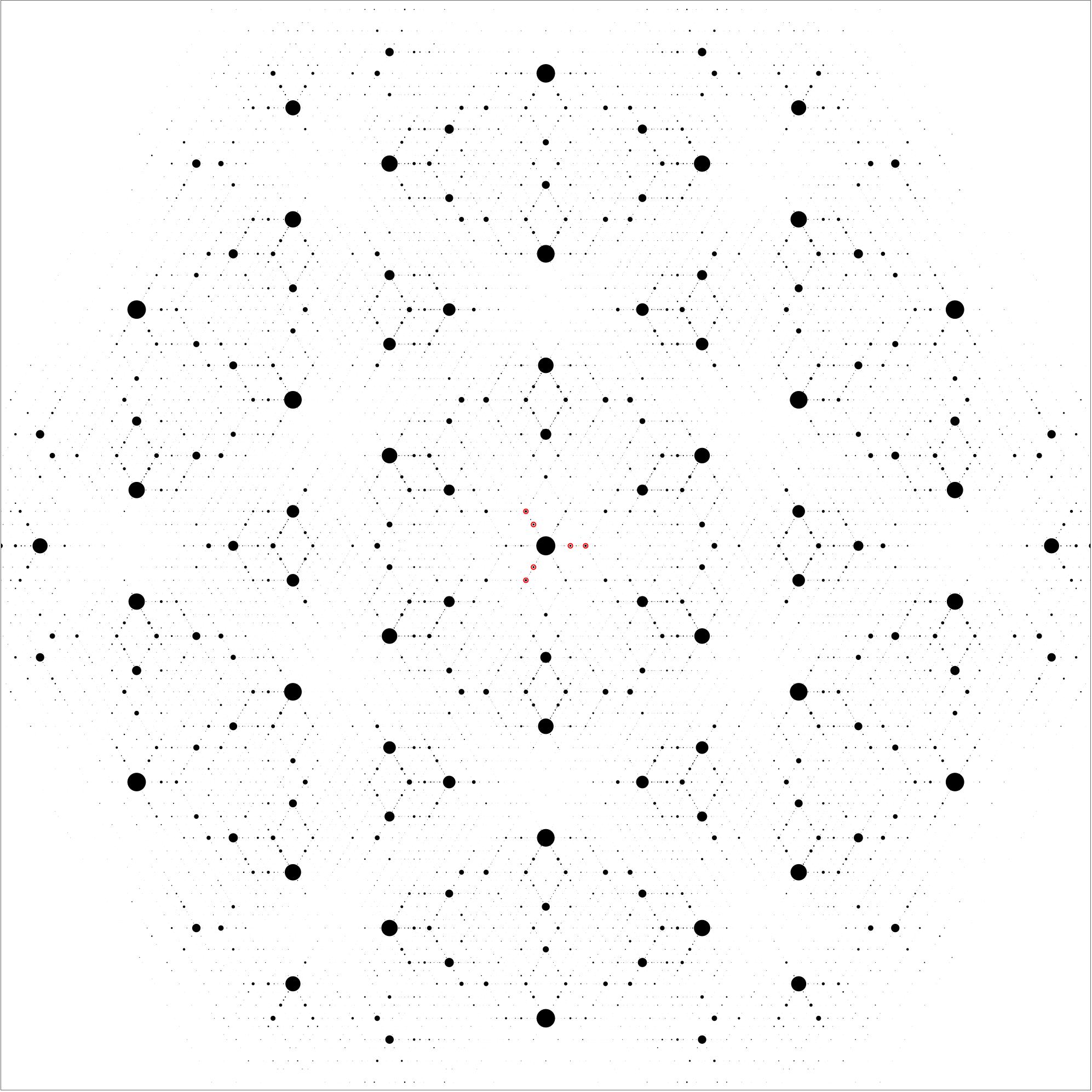}
	\caption{Fourier spectrum of the hexagonal $\Hhh$ Fibonacci tiling of Fig.~\ref{fig:Hexa55}. Disk areas are proportional to Bragg peak intensities. Red circles mark the positions of the grid vectors. 
	\label{Fig:FT}}
\end{figure*}

Finally, as shown by \citet[Eq.~(5.5)]{Rabson88}, the Ho condition~\eqref{Eq:Ho} ensures that if $\rho(\rv)$ is a sum of delta functions, located at the vertices $\vv=\sum_{j} n_j \av{j}$ of the tiling $\CT$, determined by the dual grid method, then its Fourier transform $\rho(\kv)$ is non-vanishing at most on the reciprocal lattice of integral linear combinations of the grid vectors $\kv^{(j)}$. This allows us to obtain the Fourier transform numerically by calculating the sums 
\begin{equation}\label{Eq:FT}
    \rho(\kv) 
    = \int \dd\rv e^{-i\kv\cdot\rv}\sum_{\vv\in\CT} \delta(\rv-\vv)
    = \sum_{\vv\in\CT} e^{-i\kv\cdot\vv},
\end{equation}
taking $\kv=\sum_{j} m_j \kv^{(j)}$, with the $m_j$ spanning a sufficiently large range so as to capture all Bragg peaks above a certain threshold in the region of interest in Fourier space. Figure~\ref{Fig:FT} shows a typical example, calculated for the hexagonal $\Hhh$ Fibonacci tiling of Fig.~\ref{fig:Hexa55}. The area of each disk is proportional to the Bragg peak intensity, normalized by the intensity of the Bragg peak at $\kv=0$. Red circles mark the positions of the grid vectors. The calculation was performed by summing Eq.~\eqref{Eq:FT} over more than 34000 tiling vertices, and taking $-5 \leq m_j \leq 5$.


\section{Interpretation as a projection from 6-dimensional superspace}
\label{sec:projection}




Although the dual grid method is probably the most direct approach for generating our family of tilings, it is beneficial to consider them also as projections from an abstract 6-dimensional \emph{superspace} or hyperspace. 
Accordingly, we view the six 2-dimensional tiling vectors of Eq.~\eqref{Eq:a-star} as two orthogonal 6-dimensional vectors $(a_\mu^{(1)},\ldots, a_\mu^{(6)})$ [$\mu=1,2$], spanning the 2-dimensional physical subspace $\CE$ of a 6-dimensional superspace. We then need to choose four additional 6-dimensional vectors $(B_\mu^{(1)},\ldots, B_\mu^{(6)})$ [$\mu=3,4,5,6$], orthogonal to the first two, to span the orthogonal complement of $\CE$, which is a 4-dimensional \emph{internal} or \emph{perpendicular} space $\CE^\perp$. 

The choice of these four vectors is not unique, but once they are chosen everything else is fixed. In particular, this process adds 4 components to each of the original tiling vectors, extending them into the six 6-dimensional orthogonal vectors $\AV{j} = (\av{j},\BV{j})$ that also span the entire superspace. This defines the so-called \emph{star map}, associating with every potential 2-dimensional vertex $\vv$ a unique 4-dimensional orthogonal complement, $\vv^\star$, given by

\begin{equation}\label{Eq:StarMap}
    \star:\quad \vv=\sum_{j} n_j \av{j}
    \quad \longrightarrow \quad
    \vv^\star=\sum_{j} n_j \BV{j}.
\end{equation}

In making our choice, we note that the first three superspace basis vectors, $\AV{1}$, $\AV{2}$, and $\AV{3}$, generate a 3-dimensional cubic lattice and project along its body diagonal onto the short grid vectors $\av{1}$, $\av{2}$, and $\av{3}$. The remaining three, $\AV{4}$, $\AV{5}$, and $\AV{6}$, form a second 3-dimensional cubic lattice, orthogonal to the first, and project along its body diagonal onto the long grid vectors $\av{4}$, $\av{5}$, and $\av{6}$. The two body-diagonals are therefore orthogonal to physical space $\CE$, and can be taken as two of the four basis vectors that generate the 4-dimensional internal space $\CE^\perp$. The remaining 2-dimensional subspace $\CE'$ of internal space is obtained by viewing the three pairs of vectors, $\AV{j}$ and $\AV{j+3}$ for $j=1,2,3$, as generating three 2-dimensional square lattices, each of which can be used to generate a 1-dimensional Fibonacci tiling, using the standard projection scheme (see \citet[Fig 2.9]{Senechal96}). This is done by applying a $90^\circ$ rotation to the corresponding projections onto $\CE$, $(a_\mu^{(j)},a_\mu^{(j+3)})$ [$\mu=1,2$], to obtain the orthogonal vectors $(b_\mu^{(j)}, b_\mu^{(j+3)}) = (a_\mu^{(j+3)}, -a_\mu^{(j)})$, and construct the 2-dimensional projections onto $\CE'$, given by
\begin{equation}\label{Eq:b-star}
    \bv{j}=
    \begin{cases}
    \av{j+3} &j=1,2,3,\\
    -\av{j-3} &j=4,5,6,
    \end{cases}
\end{equation}
as shown in Fig.~\ref{fig:bvecs}. Thus, each tiling vector $\av{j}$ is extended by a 4-dimensional internal-space vector $\BV{j}=(\bv{j}, c_s^{(j)}, c_l^{(j)}) \in \CE^\perp$, whose components include a 2-dimensional internal-space vector $\bv{j} \in \CE'$, and two additional components $c_s^{(j)}$ and $c_l^{(j)}$ along the body diagonals.

For convenience, our choice for the embedding of physical space into superspace is summarized by the matrix
\begin{eqnarray}\label{Eq:AMatrix}
    \hskip-10pt
    \renewcommand*{\arraystretch}{1.7}
    \frac{A}{(\frac{2\tau}{3\sqrt{5}})} = 
    \begin{pmatrix}
    \frac{1}{\tau} &-\frac{1}{2\tau} &-\frac{1}{2\tau} &1 &-\frac{1}{2} &-\frac{1}{2}\\
    0 &\frac{\sqrt{3}}{2\tau} &-\frac{\sqrt{3}}{2\tau} &0 &\frac{\sqrt{3}}{2} &-\frac{\sqrt{3}}{2}\\
     1 &-\frac{1}{2} &-\frac{1}{2} &-\frac{1}{\tau} &\frac{1}{2\tau} &\frac{1}{2\tau}\\
     0 &\frac{\sqrt{3}}{2} &-\frac{\sqrt{3}}{2} &0 &-\frac{\sqrt{3}}{2\tau} &\frac{\sqrt{3}}{2\tau}\\
     \sqrt{\frac{\sqrt{5}}{2\tau}} &\sqrt{\frac{\sqrt{5}}{2\tau}} &\sqrt{\frac{\sqrt{5}}{2\tau}} &0 &0 &0\\
      0 &0 &0 &\sqrt{\frac{\sqrt{5}}{2\tau}} &\sqrt{\frac{\sqrt{5}}{2\tau}} &\sqrt{\frac{\sqrt{5}}{2\tau}}
    \end{pmatrix}
\end{eqnarray}
where the $j^{th}$ column of the matrix $A$ is proportional to the components of $\AV{j}=(\av{j}, \bv{j}, c_s^{(j)}, c_l^{(j)})$. Alternatively, the first two rows of $A$ span the 2-dimensional physical space $\CE$, the third and fourth rows span the 2-dimensional internal subspace $\CE'$, and the last two rows span the two 1-dimensional internal subspaces along the two cubic body-diagonals.

According to the \emph{cut-and-project} scheme, the tiling $\CT$ contains only those vertices $\vv=\sum_{j} n_j \av{j}$ whose internal-space complement $\vv^\star$, given by the star map~\eqref{Eq:StarMap}, lies within a shifted \emph{window}, given by
\begin{equation}\label{Eq:Window}
    W+\fv 
    = \left\{\sum_{j=1}^6 \left(\lambda_j+f_j\right) \BV{j} 
    \ \Bigg{|} \  0< \lambda_j <1\right\}.
\end{equation}
This procedure generates a tiling that is indistinguishable from the one generated using the dual grid method with the same choice of grid shifts $f_j$ that define the shift vector $\fv=\sum_j f_j\BV{j}$~\cite{Rabson88}. Note that the window $W$ is the projection onto internal space $\CE^\perp$ of a single 6-dimensional hypercube. 

When projecting the vertices of the 6-dimensional hypercubic lattice onto the different subspaces, the 2-dimensional spaces $\CE$ and $\CE'$ become densely filled with projected points, but the 1-dimensional subspaces each contain a periodic lattice of projected points. To within a scale factor, a particular vertex $\VV=\sum_{j} n_j \AV{j}$ is projected onto a point at height $h_s=n_1+n_2+n_3$ on the first cubic body-diagonal, and a point at height $h_l=n_4+n_5+n_6$ on the second cubic body-diagonal. Accordingly, the 4-dimensional window is not densely filled with projected points, but rather contains 2-dimensional cross sections, determined by the pair of integer heights, or levels, $(h_s,h_l)$. A similar situation occurs in the projection that generates the dodecagonal tiling, as described by \citet{Socolar89}.

By projecting the shifted window~\eqref{Eq:Window} onto the two cubic body-diagonals, we can obtain the first set of conditions for a vertex $\VV=\sum_{j} n_j \AV{j}$ to be projected into the window, namely
\begin{subequations}\label{Eq:Height}
\begin{equation}\label{Eq:Height-s}
    \alpha_s < h_s = \sum_{j=1}^3 n_j < 3+\alpha_s,
\end{equation}
and
\begin{equation}\label{Eq:Height-l}
    \alpha_l < h_l = \sum_{j=4}^6 n_j < 3+\alpha_l,
\end{equation}
\end{subequations}
where we have invoked the definition of the structure invariants of Eq.~\eqref{Eq:invariants}. There are, in general, nine pairs of height combinations as $h_s,h_l\in\{1,2,3\}$, unless one or both of the structure invariants are zero, in which case the number of height combinations reduces to six or four, respectively. This is because according to Eqs.~\eqref{Eq:Height}, if a structure invariant is zero, the corresponding height can take only the values 1 or 2.

\begin{figure*}
    \centering
    \begin{subfigure}[b]{0.49\textwidth}
      \centering
      \includegraphics[width=\linewidth]{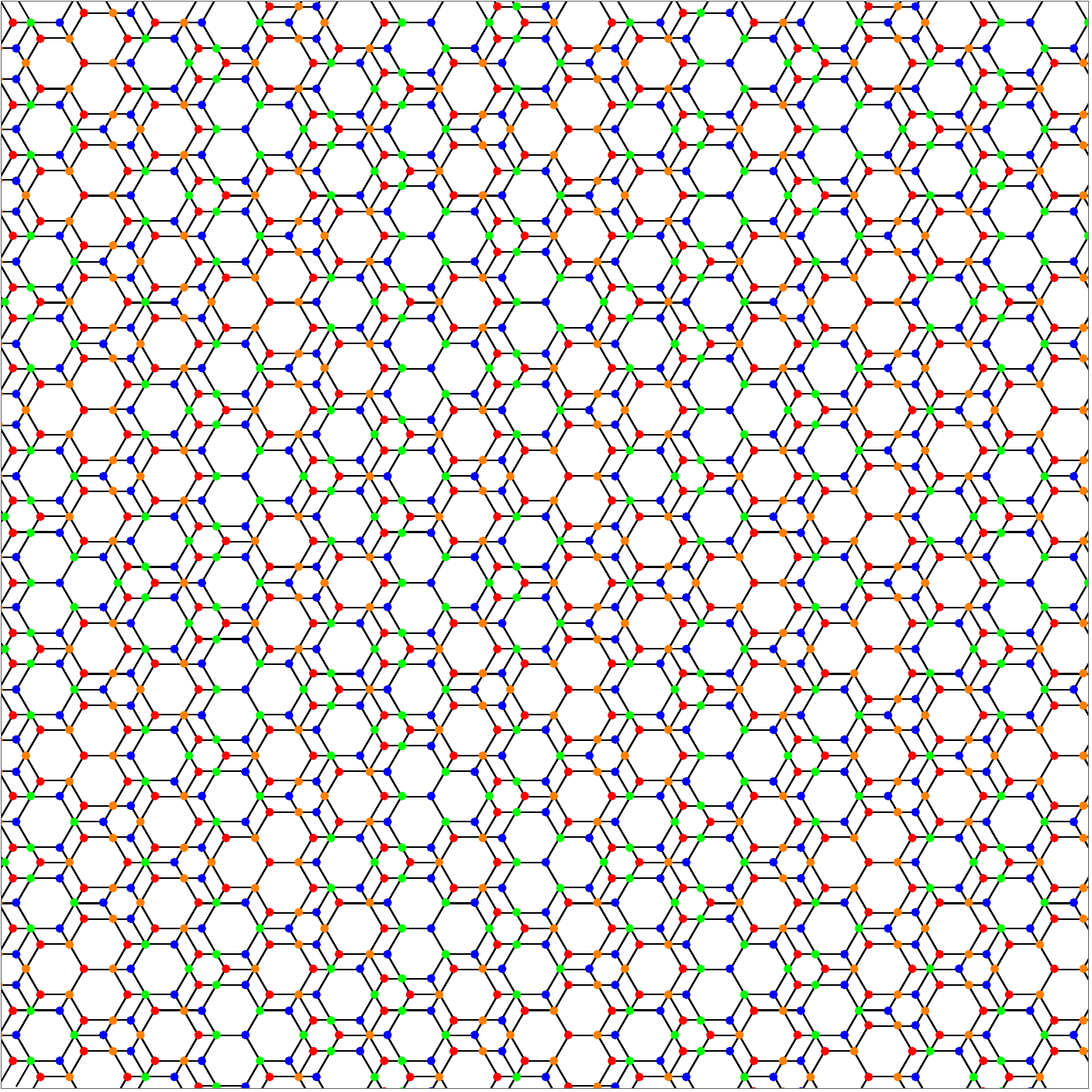}
      \makebox[8.5pt][r]{
        \raisebox{151.8pt}{%
          \includegraphics[width=.35\linewidth]{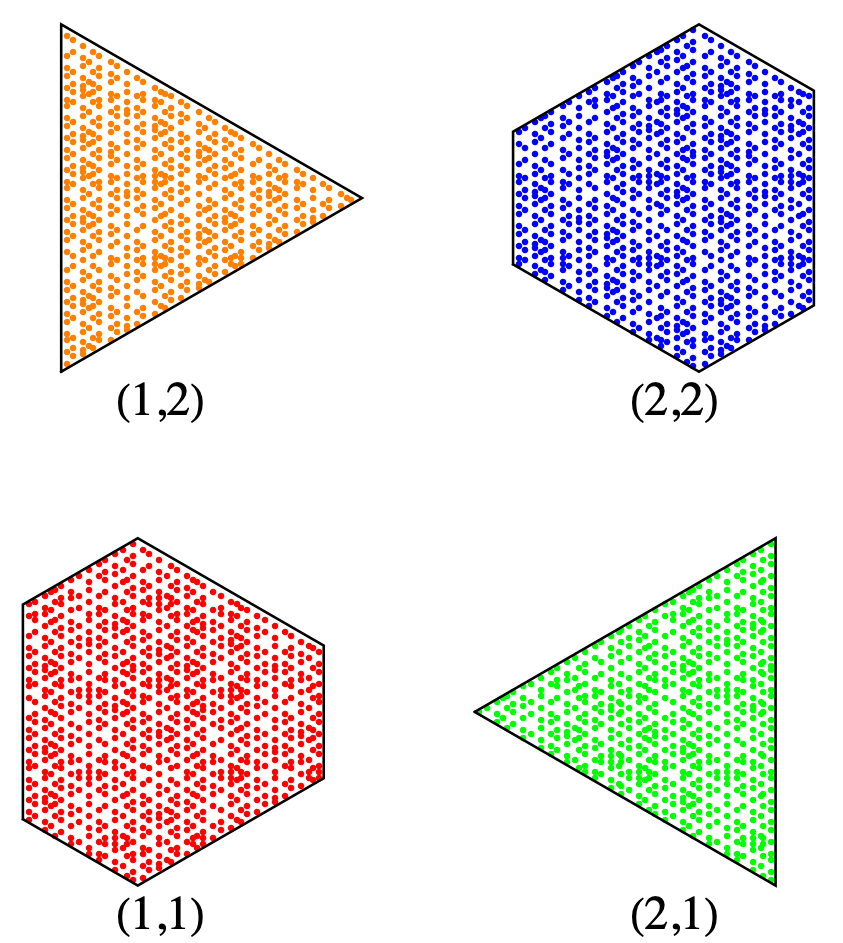}
        }\hspace*{1em}%
        }
    \caption{Hexagonal $\Hzz$ tiling with $\alpha_s\equiv\alpha_l\equiv0$ as in Fig.~\ref{fig:Hexa00}.}
    \label{fig:heights00}
    \end{subfigure}
    \begin{subfigure}[b]{0.49\textwidth}
      \centering
      \includegraphics[width=\linewidth]{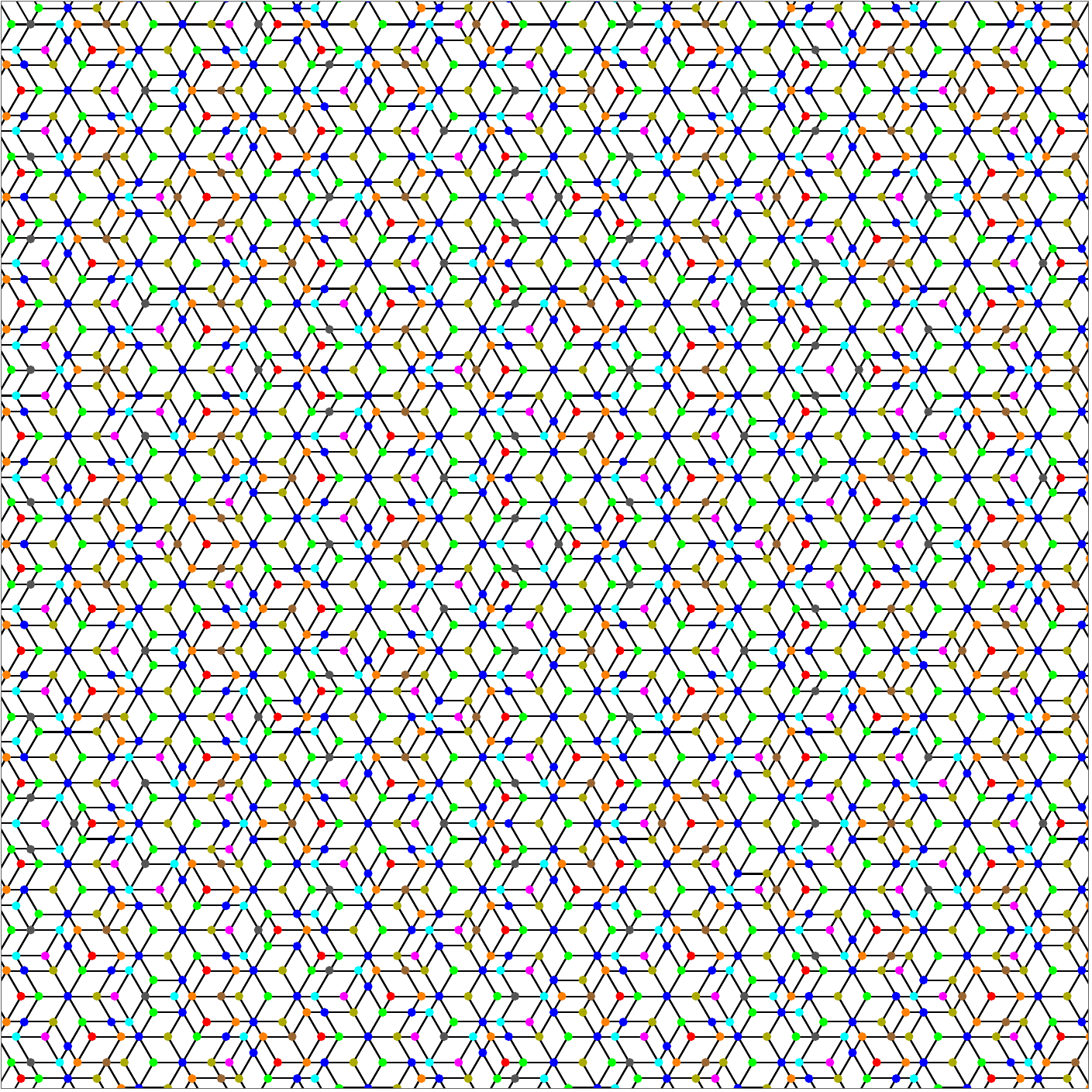}
      \makebox[8.3pt][r]{
        \raisebox{107.8pt}{%
          \includegraphics[width=.5\linewidth]{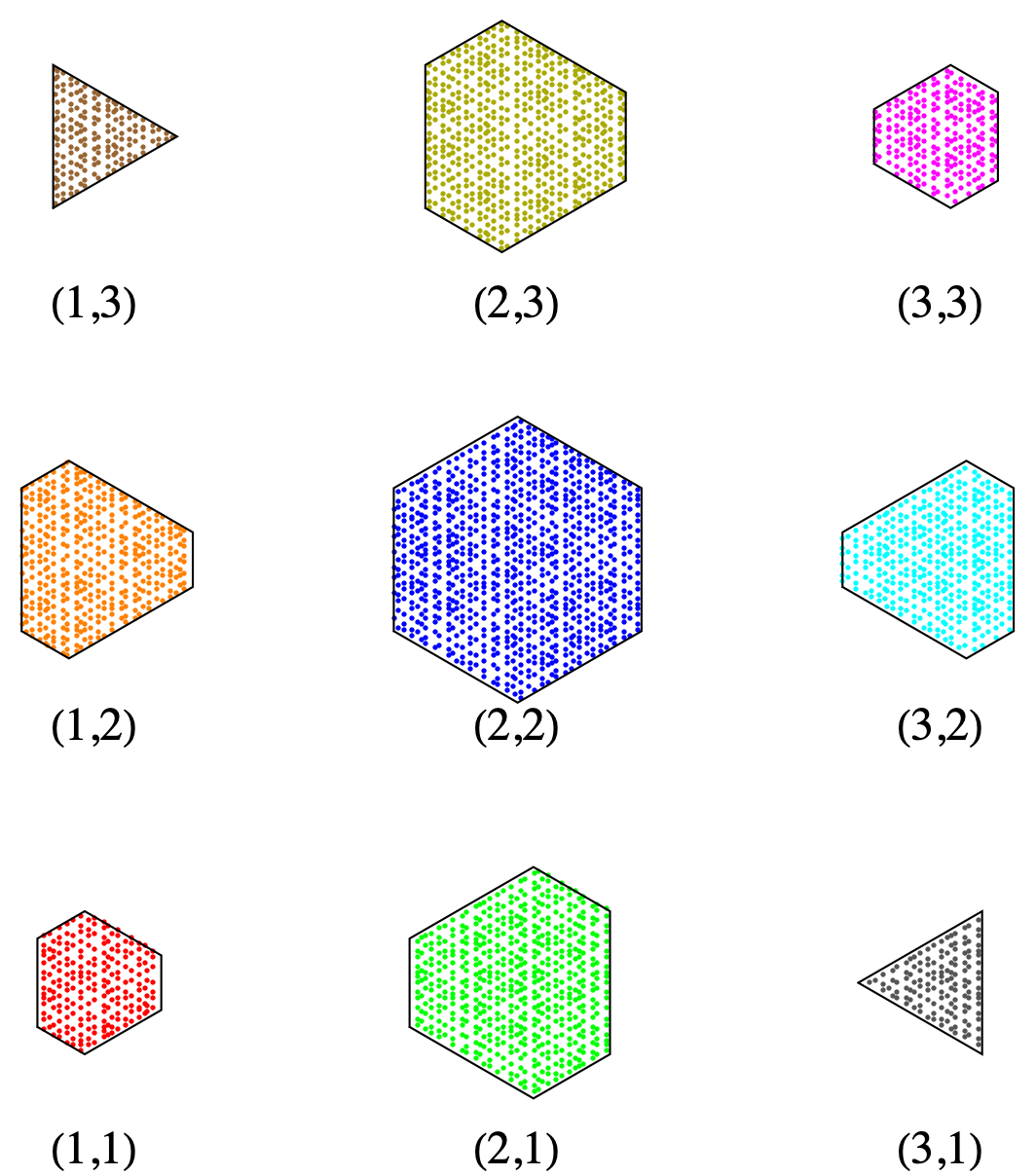}
        }\hspace*{1em}%
        }
    \caption{Hexagonal $\Hhh$ tiling with $\alpha_s\equiv\alpha_l\equiv0.5$ as in Fig.~\ref{fig:Hexa55}.}
    \label{fig:heights0505}
    \end{subfigure}
    \vskip3pt
    \begin{subfigure}[b]{0.49\textwidth}
      \centering
      \includegraphics[width=\linewidth]{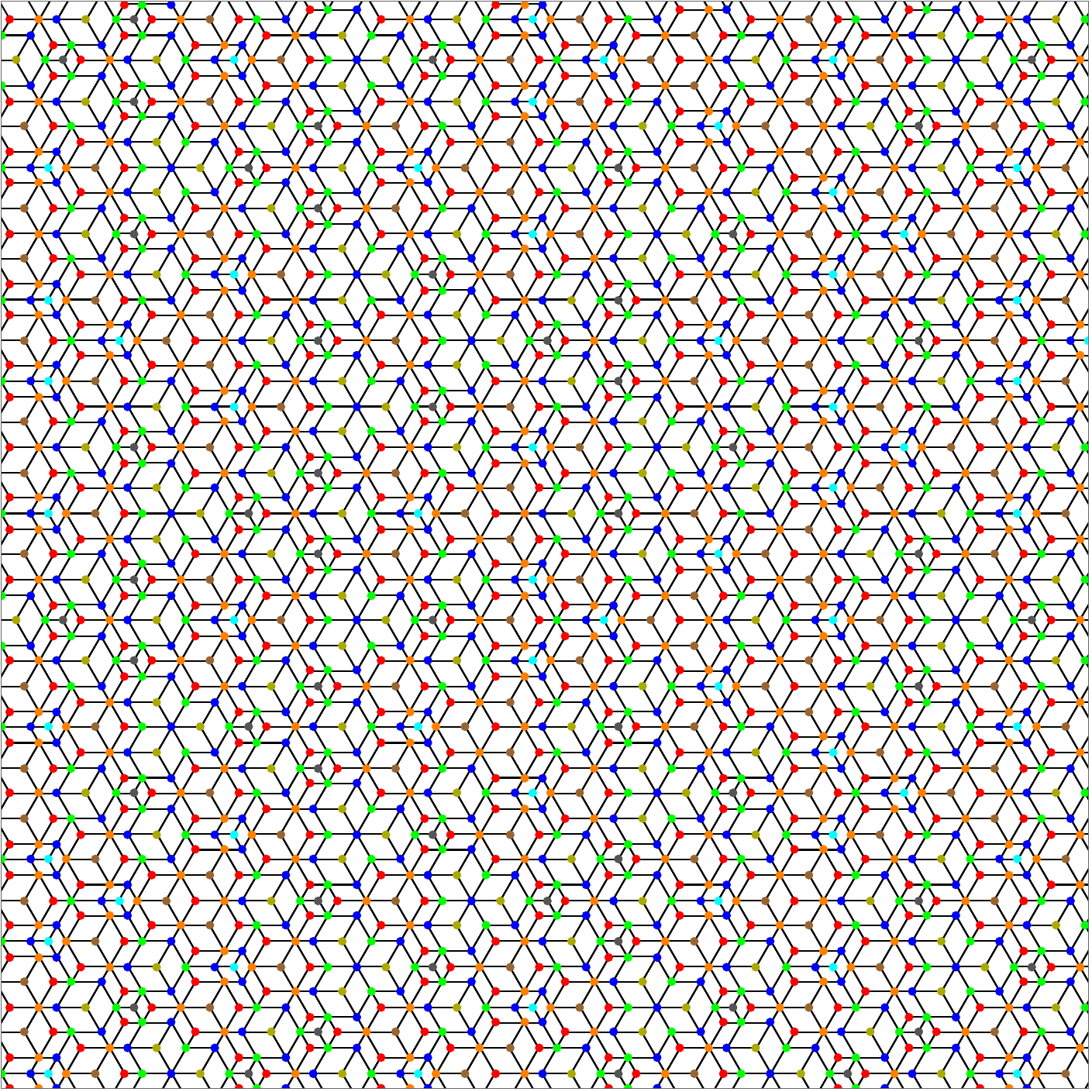}
      \makebox[8.3pt][r]{
        \raisebox{115pt}{%
          \includegraphics[width=.5\linewidth]{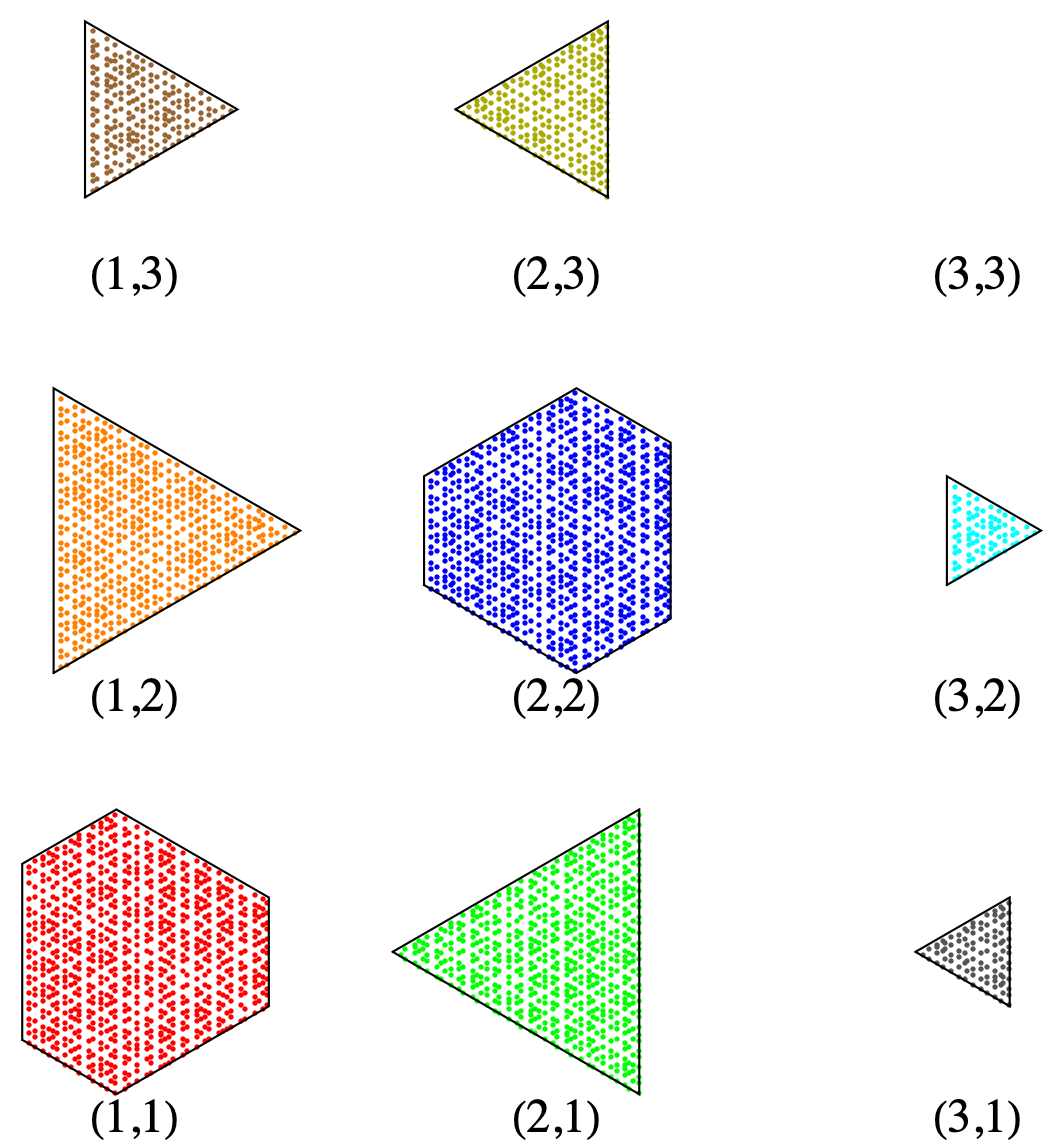}
        }\hspace*{1em}%
        }
    \caption{Trigonal tiling with $\alpha_s\equiv\alpha_l\equiv10^{-8}$ as in Fig.~\ref{fig:Trig00}.}
    \label{fig:heights10-8}
    \end{subfigure}
    \begin{subfigure}[b]{0.49\textwidth}
      \centering
      \includegraphics[width=\linewidth]{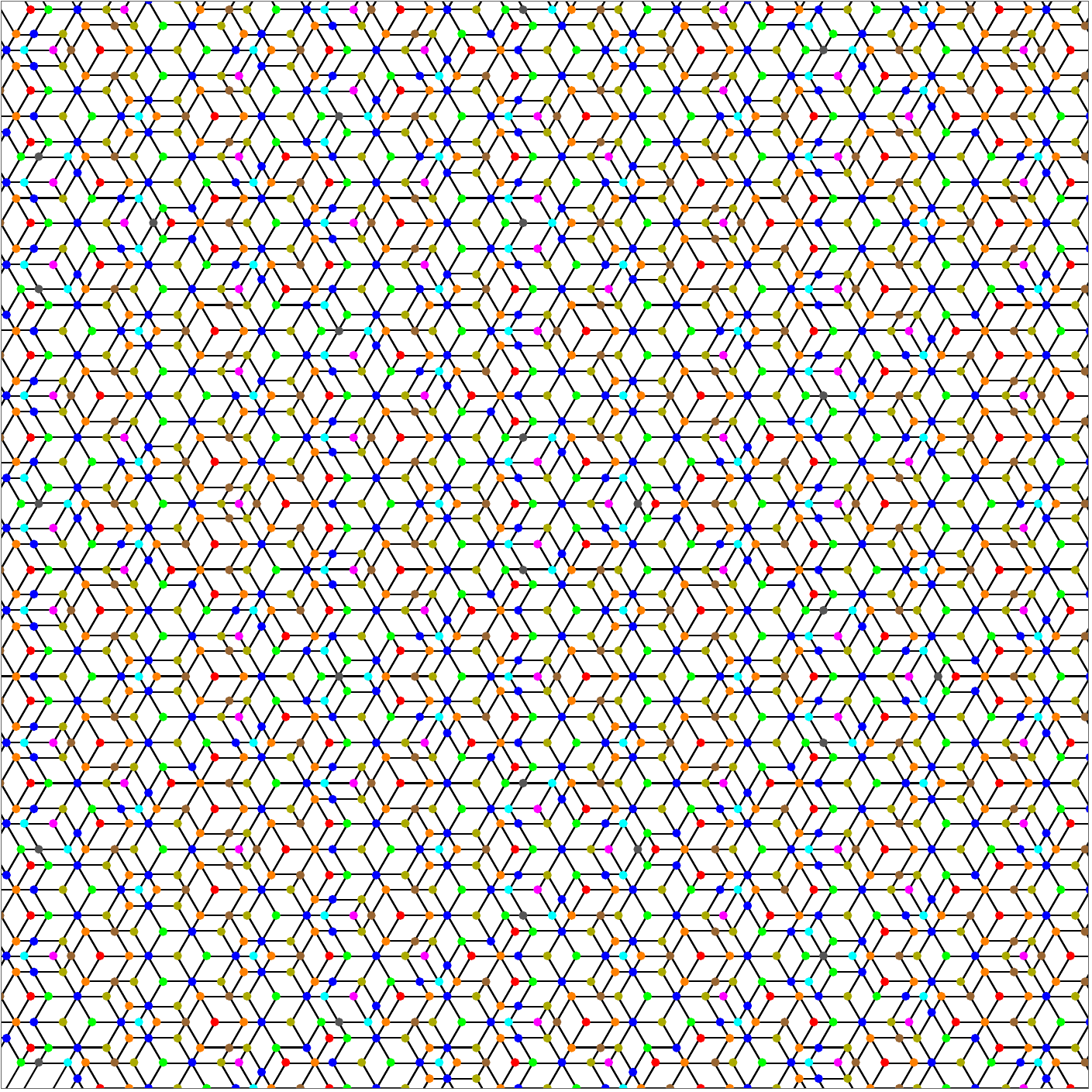}
      \makebox[8.3pt][r]{
        \raisebox{106.7pt}{%
          \includegraphics[width=.5\linewidth]{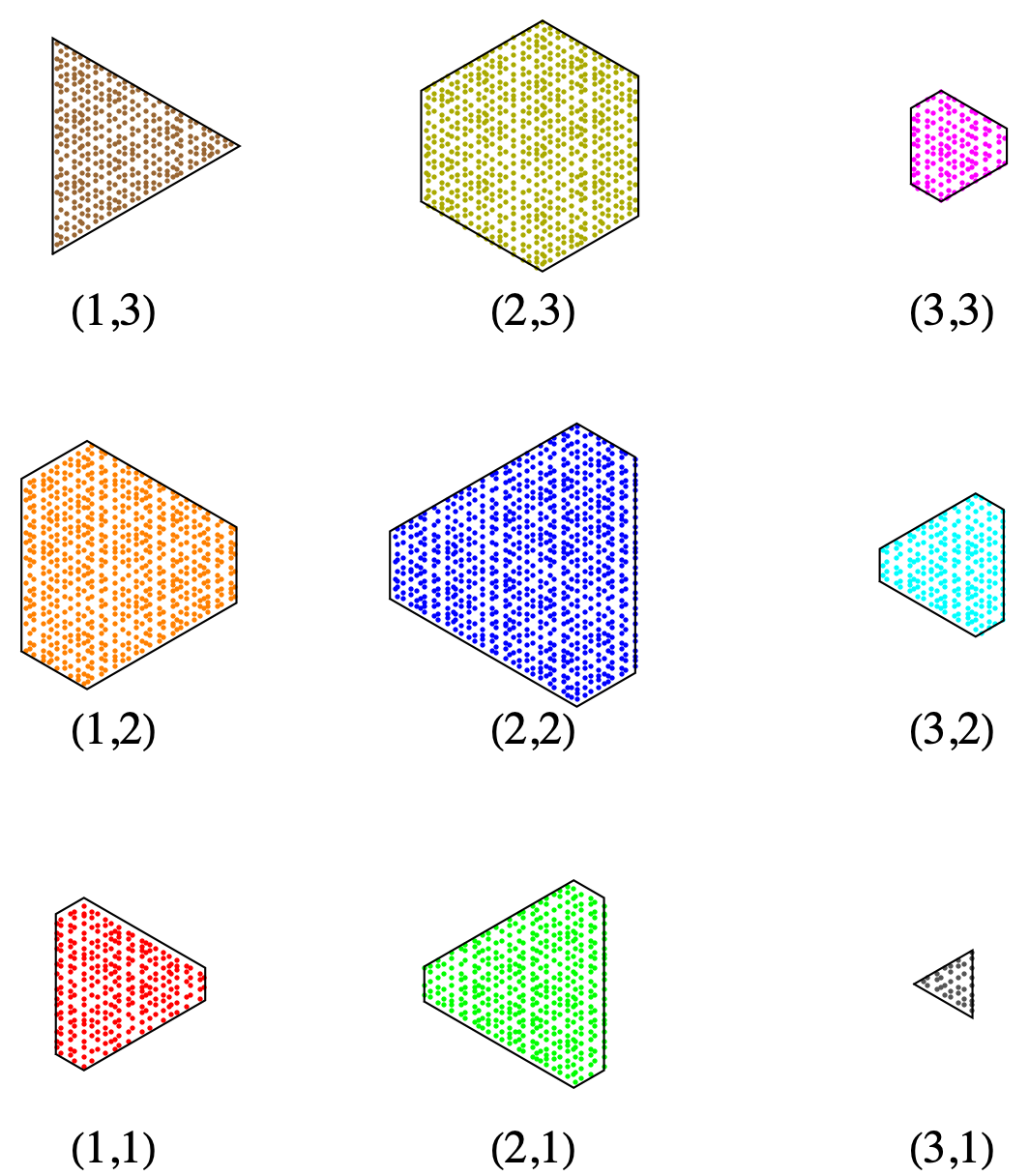}
        }\hspace*{1em}%
        }
    \caption{Trigonal tiling with $\alpha_s\equiv0.2$ and $\alpha_l\equiv0.7$ as in Fig.~\ref{fig:Trig27}.}
    \label{fig:heights27}
    \end{subfigure}
\caption{Hexagonal and trigonal Fibonacci tilings  generated using the cut-and-project method. The vertices of all tilings are color coded according to their level indices $(h_s,h_l)$, corresponding to their projections onto internal space, shown in the inset. The inset shows the 2-dimensional cross sections through the window corresponding to the different height combinations $(h_s,h_l)$, defined in Eqs.~\eqref{Eq:Height}. (a) and (b) Hexagonal tilings of Figs.~\ref{fig:Hexa00} and \ref{fig:Hexa55}; (c) and (d) Trigonal tilings of Figs.~\ref{fig:Trig00} and \ref{fig:Trig27}. See the text for a complete discussion.
\label{Fig:VertexHeights}}
\end{figure*}

For each possible level $(h_s,h_l)$, we obtain the corresponding shape of the 2-dimensional cross section of the window, in two steps. First we cut each 3-dimensional cube normal to its diagonal, a distance $h-\alpha$ from its bottom corner. This gives us a region which we label $\CR(h,\alpha)$. It has the shape of a triangle if we cut the cube in its lower third, a triangle pointing in the opposite direction in the upper third of the cube, and a hexagon in the middle third, which is regular only when it contains the body center. The total cross section of the window is then given by the pairwise sum of points from the two individual cross sections, also known as their Minkowski sum. Thus, the third condition for a vertex 
$\VV=\sum_{j} n_j \AV{j}$ to be projected into the window, and therefore included in the tiling is
\begin{align}\label{Eq:b-window}\nonumber
    &\sum_{j} n_j \bv{j} \in \CR(h_s,\alpha_s)+\CR(h_l,\alpha_l)\\
    &= \left\{ {\bf b}_s+{\bf b}_l\ \bigg{|}\ {\bf b}_s\in \CR(h_s,\alpha_s), {\bf b}_l\in \CR(h_l,\alpha_l)\right\}.
\end{align}

Figure~\ref{Fig:VertexHeights} shows a few examples of hexagonal and trigonal Fibonacci tilings generated using the cut-and-project method. The tiling vertices are color-coded according to their level indices $(h_s,h_l)$. Their images, as they fall into the cross sections of the window under the star map~\eqref{Eq:StarMap}, are shown in the insets. Because the vertices fill the cross sections uniformly, the frequency in which each vertex level appears in the tiling is proportional to the area of the corresponding cross section. 

In the hexagonal tilings in the upper row, every cross section that has only trigonal symmetry appears again, at a different level, rotated by $60^\circ$, thus maintaining the overall hexagonal symmetry of the tilings. In the trigonal tiling of Fig.~\ref{fig:heights10-8}, with $\alpha_s\equiv\alpha_l\equiv10^{-8}$, both 3-dimensional cubes are shifted ever so slightly such that the $3^{rd}$ levels barely cut through their top corners. Consequently, level (3,3) is negligibly small. It cannot be seen in the inset, and barely contributes any vertices to the tiling. The cross sections at levels ($i$,3) and (3,$i$), $i=1,2$, which are approximately the Minkowski sum of a cut at level $i$ of one cube with a point, are sufficiently large to contribute vertices to tiling. These vertices are exactly at the centers of the hexagons in Fig.~\ref{fig:heights00}. Their positions do not break the 6-fold symmetry of the hexagonal tiles. This is done by the edges connected to them, as the only possible connected vertices are one level down from ($i$,3) to ($i$,2), and from (3,$i$) to (2,$i$), thus breaking the hexagonal symmetry. In the trigonal tiling of Fig.~\ref{fig:heights27}, with $\alpha_s\equiv0.2$ and $\alpha_l\equiv0.7$, the hexagonal symmetry is strongly broken, with no pairs of cross sections that are related by a 6-fold rotation (compare with Fig.~\ref{fig:heights0505}).

As expected, the hexagonal $\Hzz$ tiling of Fig.~\ref{fig:heights00}, with $\alpha_s\equiv\alpha_l\equiv0$, has only four levels, leading to a smaller variation in the level indices of the different tiles. In this tiling the two hexagonal tiles originate from two possible level combinations each, related by a 6-fold rotation or a mirror reflection: the large hexagons have vertices with level indices given by [($h_s$,1),($h_s$,2),($h_s$,1),($h_s$,2),($h_s$,1),($h_s$,2)], $h_s=1,2$; and the small hexagons have vertices with level indices given by [(1,$h_l$),(2,$h_l$),(1,$h_l$),(2,$h_l$),(1,$h_l$),(2,$h_l$)], $h_l=1,2$. The parallelogram tiles consist of one vertex from each of the four levels, with the vertices given by [(1,1),(2,1),(2,2),(1,2)] to within mirror reflection. The hexagonal $\Hhh$ tiling of Fig.~\ref{fig:heights0505}, with $\alpha_s\equiv\alpha_l\equiv0.5$, originates from nine different levels, leading to a larger variety of tile indices, whose enumeration can be carried out in a similar manner.

\section{Use of substitution rules: Hexagonal tiling case studies}
\label{sec:hex}

Both the dual-grid method and the cut-and-project method allow us systematically to generate all the trigonal and hexagonal tilings, considered above, by specifying the values of the two structure invariants $\alpha_s$ and $\alpha_l$. When searching for substitution rules, one has to treat each tiling separately. We therefore conclude by concentrating on the substitution rules for the $\Hzz$ and $\Hhh$ hexagonal tilings only, while resorting to heuristic or trial-and-error methods for discovering these rules. We examine a few specific properties of these tilings more closely, namely, their tile and vertex frequencies, as well as the imbalance introduced when bipartitioning the tiling vertices into two subsets. We calculate the vertex frequencies independently using the substitution rules and the cut-and-project method, which allows us to support the validity of our proposed substitution rules. Further studies of the properties of individual members of our trigonal-hexagonal family of tilings, are left for future work.

\subsection{The hexagonal $\Hzz$ tiling}

\begin{figure}[bt]
    \centering
    \begin{subfigure}[b]{\linewidth}
      \centering
      \includegraphics[width=\linewidth]{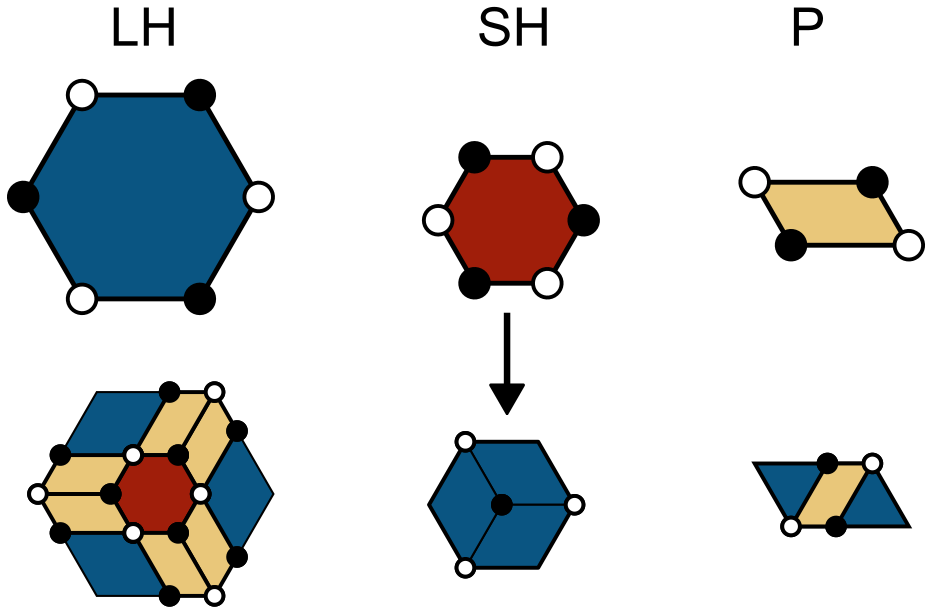}
    \caption{Substitution rules of the hexagonal $\Hzz$ tiling of Fig.~\ref{fig:Hexa00}.}
    \label{fig:deflation}
    \end{subfigure}
    \begin{subfigure}[b]{\linewidth}
      \centering
      \includegraphics[width=\linewidth]{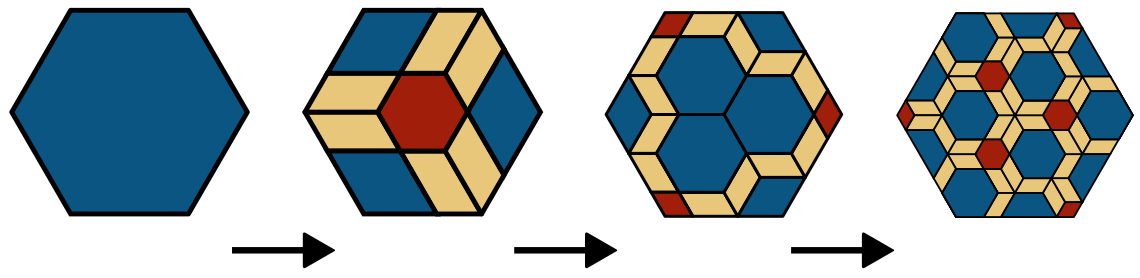}
    \caption{Three generations of substitution, starting from a single LH tile.}
    \label{fig:sequence}
    \end{subfigure}
    \caption{(a) Substitution rules for the hexagonal $\Hzz$ tiling of Fig.~\ref{fig:Hexa00}. The top row shows the three prototiles: the large hexagon (LH), the small hexagon (SH), and the parallelogram (P). Black and white circles denote the parity of the vertices: even vertices, those whose star maps of Eq.~\eqref{Eq:StarMap} lie within the hexagonal windows, shown in the inset of Fig.~\ref{fig:heights00} and in Fig.~\ref{fig:Perp00}, are colored black, while odd vertices, corresponding to the triangular windows, are white. The bottom row displays a single generation of the substitution, which depends in its orientation on the black and white coloring of the vertices. (b) Three generations of substitution, starting with a single LH tile. With every new generation the tile areas shrink by a factor of $\tau^2$.}
\end{figure}

\begin{figure}
	\centering
	\includegraphics[width=.85\linewidth]{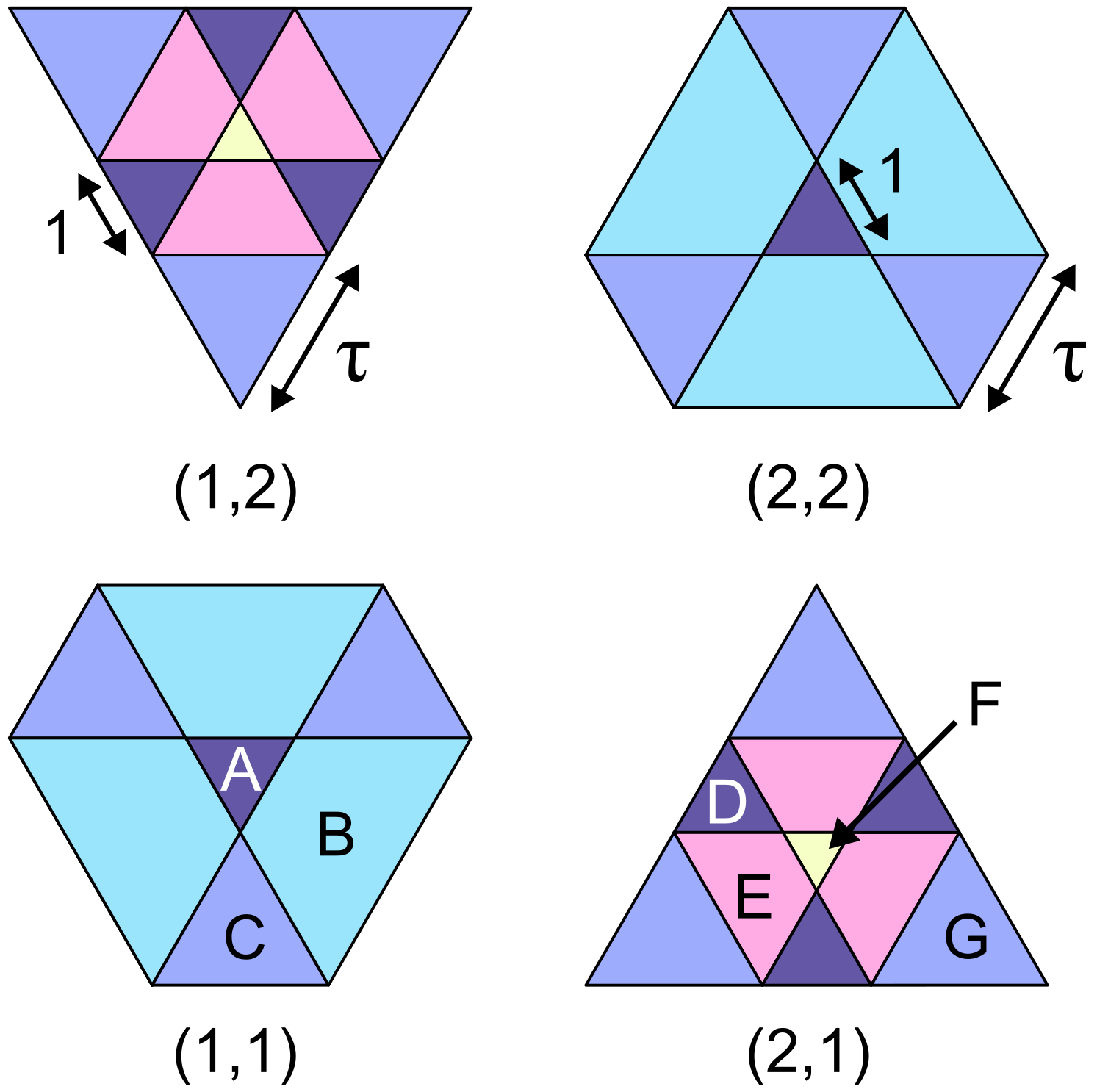}
	\caption{Two-dimensional cross sections through the internal-space window of the hexagonal $\Hzz$ tiling (as in the inset of Fig.~\ref{fig:heights00}), corresponding to the four possible level indices $(h_s,h_l)$, defined in Eqs.~\eqref{Eq:Height}. Subdomains of the cross sections are labeled according to the vertex configurations, shown in Fig.~\ref{fig:Vertices}, that they generate, and color coded according to their areas. Double arrows labeled by 1 and $\tau$ specify the relative scale of the different triangles and trapezoids. Thus, if the $A$ and $D$ triangles have unit area, then the $F$ triangle has area $\tau^{-2}$, the $C$ and $G$ triangles have area $\tau^2$, the $E$ trapezoid has area $\tau^2-\tau^{-2}=\sqrt{5}$, and the $B$ trapezoid has area $\tau^4-1=\tau^2\sqrt{5}$.
    \label{fig:Perp00} }
\end{figure}

\begin{figure}
	\centering
	\includegraphics[width=\linewidth]{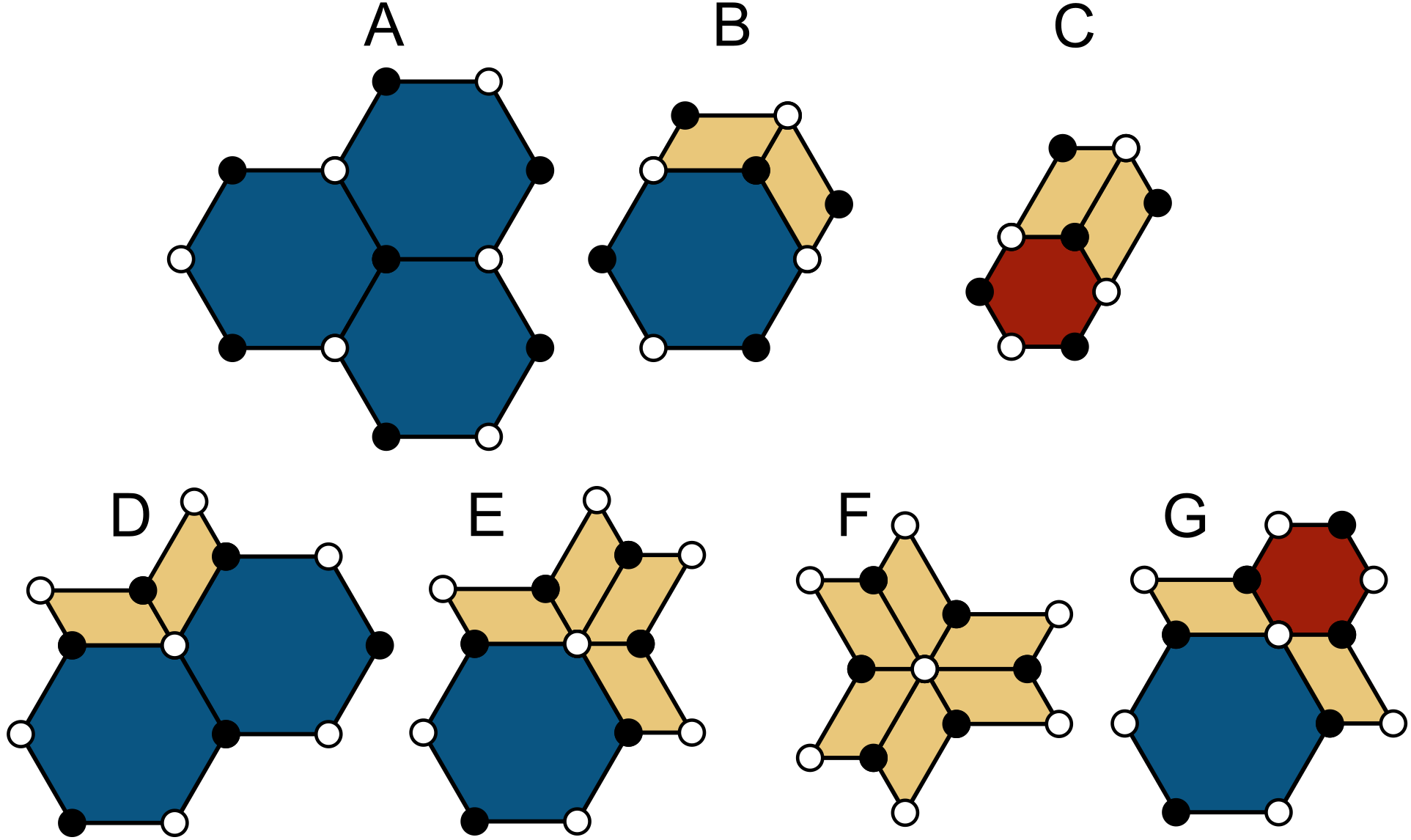}
	\caption{The seven vertex configurations of the hexagonal $H_{00}$ tiling. The configurations are split into two rows, grouped by the color (parity) of their central vertex: [A, B, C], black (even), and [D, E, F, G], white (odd), as explained in the main text. \label{fig:Vertices}}
\end{figure}

\subsubsection{Basic properties}

The hexagonal $H_{00}$ tiling of Fig.~\ref{fig:Hexa00} consists of three types of tiles, or three \emph{prototiles}---a large hexagon (LH), a small hexagon (SH), and a parallelogram (P), shown on the top of Fig.~\ref{fig:deflation}. All the long edges of these tiles are a factor of $\tau$ longer than the short ones. The vertices of the hexagonal $H_{00}$ tiling come in two varieties, depending on the parity of the sum of their level indices, corresponding to the two types of internal-space windows, shown in the inset of Fig.~\ref{fig:heights00}, and reproduced with more detail in Fig.~\ref{fig:Perp00}. The vertices associated with the hexagonal windows, colored red and blue in Fig.~\ref{fig:heights00}, have even parity and are decorated by black circles in Fig.~\ref{fig:deflation}. The remaining vertices, associated with the triangular windows, colored orange and green in Fig.~\ref{fig:heights00}, have odd parity and are are decorated by white circles in Fig.~\ref{fig:deflation}. The frequencies of the two vertex types are therefore proportional to the areas $A_\triangle = \tau^6$ and $A_{\mhexagon} = \left(\tau^8 - 3\tau^4\right)$, given in arbitrary units, of the corresponding triangular and hexagonal windows, respectively. These areas can directly be inferred from Fig.~\ref{fig:Perp00}. Thus, noting that $A_\triangle + A_{\mhexagon} = 4\tau^5$, the fractions of black and white vertices are given by 
\begin{subequations}\label{Eq:Frequency-HexTri}
\begin{align}
    f_\text{black} &= \frac{A_{\mhexagon}}{A_\triangle+A_{\mhexagon}}=\frac{3-\tau^{-1}}{4}\simeq 0.5955.\label{Eq:fblack}\\
    f_\text{white} &= \frac{A_\triangle}{A_\triangle+A_{\mhexagon}}=\frac{\tau}{4}\simeq 0.4045, \label{Eq:fwhite}
\end{align}
\end{subequations}

We note in passing that each tile edge connects a black vertex to a white vertex, without any frustration, because all three tiles have an even number of edges. Consequently, the black and white vertex coloring partitions the vertices into two sets, with no edges connecting vertices within a given set. The bipartition that is generated in this way is imbalanced, and therefore not symmetric under the exchange of black and white, as there are more black vertices than white ones. This is in contrast to the familiar bipartitioning of a square lattice or the octagonal Ammann-Beenker tiling, which is symmetric under the exchange of black and white, allowing for antiferromagnetic long-range order~\cite{colorsym,Lifshitz22}. 

The $\Hzz$ tiling has seven distinct vertex configurations, corresponding to the different triangular and trapezoidal subdomains of the two types of internal-space windows, labeled as $A$ though $G$ in Fig.~\ref{fig:Perp00}. Figure \ref{fig:Vertices} shows the allowed vertex configurations themselves. The top row shows the vertex configurations that arise from the subdomains [A, B, C] of the hexagonal window and are therefore centred on black vertices, while the bottom row shows the vertex configurations that arise from the subdomains [D, E, F, G] of the triangular window and are centred on white vertices. Adding the total areas of the seven different subdomains, and dividing by the total area $A_\triangle + A_{\mhexagon} = 4\tau^5$, immediately yields the frequencies of the seven vertex configurations,
\begin{equation}\label{Eq:00VerFreq}
\begin{aligned}
&f(\text{A}) = \frac{1}{4\tau^5} \simeq 0.0225;\quad
&f(\text{B}) = \frac{3 \sqrt{5}}{4\tau^3} \simeq 0.3959;\\[0.5em]
&f(\text{C}) = \frac{3}{4\tau^3} \simeq 0.1771;\quad
&f(\text{D}) = \frac{3}{4\tau^5} \simeq 0.0676;\\[0.5em]
&f(\text{E}) = \frac{3 \sqrt{5}}{4\tau^5} \simeq 0.1512;\quad
&f(\text{F}) = \frac{1}{4\tau^7} \simeq 0.0086;\\[0.5em]
&f(\text{G}) = \frac{3}{4\tau^3} \simeq 0.1771. &\\
\end{aligned}
\end{equation}

\begin{figure*}
\centering
\includegraphics[width=.95\linewidth]{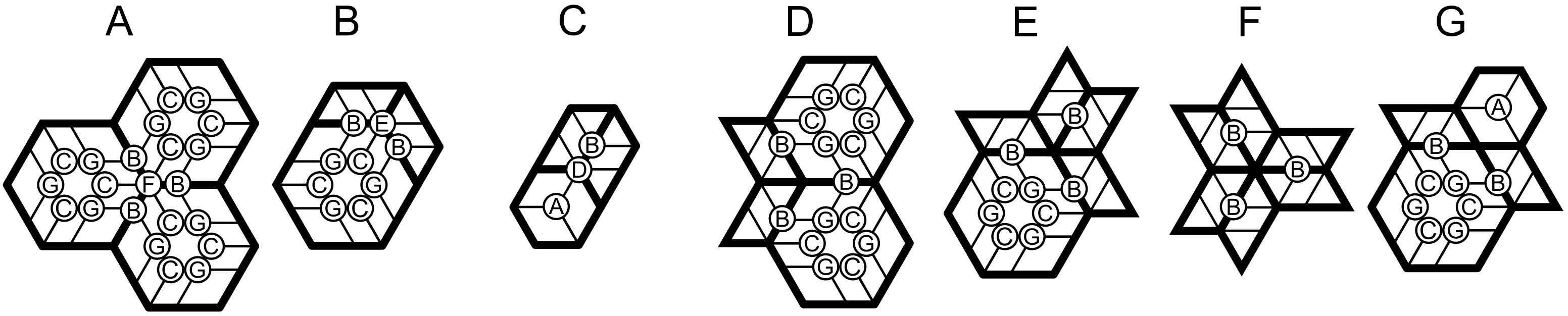}
\caption{Substitution of the seven vertex configurations of the $\Hzz$ tiling, shown in Fig.~\ref{fig:Vertices}. The current generation tiles, shown with thick black edges, are substituted into next-generation tiles with thin black edges, creating next-generation vertices. All the next-generation vertices, which are to be counted in the substitution of the original configuration, are labelled by their letters.} 
\label{fig:vertex_deflation}
\end{figure*}

\subsubsection{Substitution rules}

The proposed substitution rules for the three different tiles of the hexagonal $\Hzz$ tiling, obtained through a process of trial-and-error, are shown at the bottom of Fig.~\ref{fig:deflation}. Note that the rules for the two hexagonal tiles depend on their orientation on the black and white coloring of the vertices. Figure \ref{fig:sequence} shows three generations of substitution, starting from a single LH tile. Multiple substitutions eventually lead to a tiling that looks very much like the $\Hzz$ tiling, shown in Fig.~\ref{fig:Hexa00}. Lacking a rigorous proof that our proposed substitution rules indeed generate the $\Hzz$ tiling, we show below that a recalculation of the vertex configurations, based on the properties of the substitution, yields the exact same results as those given in Eq.~\eqref{Eq:00VerFreq}. This provides compelling evidence that the proposed substitution rules are very likely the correct ones. 

Before doing so, let us first use the substitution rules to perform a textbook calculation of the tile frequencies~\cite{Senechal96}. To do so, we consider the substitution matrix,
\begin{equation}\label{Eq:M00}
M_{00} = 
\kbordermatrix{
	& LH & SH & P\\
	LH & 1  & 1 & \frac{1}{3} \\[0.2em]
	SH & 1  & 0 & 0  \\
	P &  6  &  0 & 1 
},
\end{equation}
whose columns indicate how many small, or deflated, tiles of each type are contained within a given substituted tile. For example, the leftmost column indicates that the LH tile is substituted into a single deflated LH tile (the fact that it is cut into three pieces is unimportant), a single SH tile, and six P tiles, as shown in Fig.~\ref{fig:deflation}. Under mild conditions, regardless of the starting tile, repeated application of a substitution matrix yields tile frequencies that are given by the components of the eigenvector, corresponding to the largest eigenvalue of the matrix (for details see~\cite{Senechal96}). In the case of $M_{00}$, the largest eigenvalue is $\tau^2$, and its corresponding eigenvector is $(\tau^2, 1, 6\tau)$. Thus, the SH tile is the least frequent; LH appears $\tau^2=\tau+1$ more often; and the P tile, which is the most frequent, appears $6\tau$ more often than the SH tile. These tile frequencies are consistent with an empirical inspection of the tiling in Fig.~\ref{fig:Hexa00}.


A similar, yet slightly less commonly used procedure~\cite{Kumar86,Zobetz92}, can be applied to obtain analytical values for the frequencies of the distinct vertex configurations of Fig.~\ref{fig:Vertices}. Instead of writing a substitution matrix for tiles, one writes a substitution matrix for vertices. As in the case of tiles, under mild assumptions, regardless of the starting vertex configuration, repeated substitutions lead to a distribution of vertex configurations that is determined by the components of the eigenvector, corresponding to the largest eigenvalue of the vertex substitution-matrix. One only needs to be careful not to over count next-generation vertices that are shared with neighboring vertex configurations of the current generation.

To be precise, each vertex configuration, labeled $A$ to $G$ in Fig.~\ref{fig:Vertices}, consists of the set of tiles that surround a given vertex. In each generation of the substitution, new vertices appear in place of the original one, as shown in Fig.~\ref{fig:vertex_deflation}. If a next-generation vertex appears at the exact position of the original one it is not shared with any other vertex of the current generation, and should be counted in full. If a next-generation vertex appears on an edge between two tiles of the current-generation configuration, this next-generation vertex will be shared with a second current-generation vertex located on the opposite end of the edge, and will be counted twice unless its contribution to each current-generation vertex is halved. Finally, if a next-generation vertex appears within a tile of the original configuration, it will be shared with all the $N$ current-generation vertices around the tile, and therefore its contribution to each of these $N$ vertices must be divided by $N$, to avoid over counting. 

For example, the leftmost vertex configuration in Fig.~\ref{fig:vertex_deflation}, labeled $A$, consisting of three LH tiles, is replaced with: (1) a single F vertex, located in the same position as the original $A$ vertex, and therefore counted in full; (2) three $B$ vertices located on the edges of the large LH tiles and therefore counted as three halves; (3) three $C$ vertices in each of the three large LH tiles, each shared with all six vertices of the tile and therefore counted as three sixths in each tile; and (4) three $G$ vertices within each large LH tile, counted the same as the $C$ tiles. Altogether, in each iteration one has $A \rightarrow \frac{3}{2}B + \frac{3}{2}C + F + \frac{3}{2}G$. One thus obtains the vertex substitution-matrix
\begin{equation}\label{eq:00vertices}
\centering
V_{00}=\kbordermatrix{
	& A & B & C & D & E & F & G \\
A	& 0  & 0 & \frac{1}{6}  & 0 &0  &0 & \frac{1}{6} \\[0.3em]
B	& \frac{3}{2}   & 1 & \frac{1}{2}  & \frac{3}{2} &\frac{3}{2} & \frac{3}{2}&1\\[0.3em]
C	& \frac{3}{2}   & \frac{1}{2}  & 0  & 1 & \frac{1}{2} &0 & \frac{1}{2} \\[0.3em]
D	& 0  & 0 &  1 &  0 & 0&0 &0 \\
E	& 0  & 1 &  0 &  0 & 0&0 &0\\
F	& 1  & 0 &  0 &  0 & 0&0 &0\\
G	&  \frac{3}{2}  & \frac{1}{2}  & 0  &  1&\frac{1}{2}  &0 &\frac{1}{2} 
},
\end{equation}
whose largest eigenvalue is again $\tau^2$, and whose corresponding eigenvector gives the exact distribution of vertex configuration, as that given by Eq.~\eqref{Eq:00VerFreq}. 

\begin{figure}[b]
	\centering
	\includegraphics[width=\linewidth]{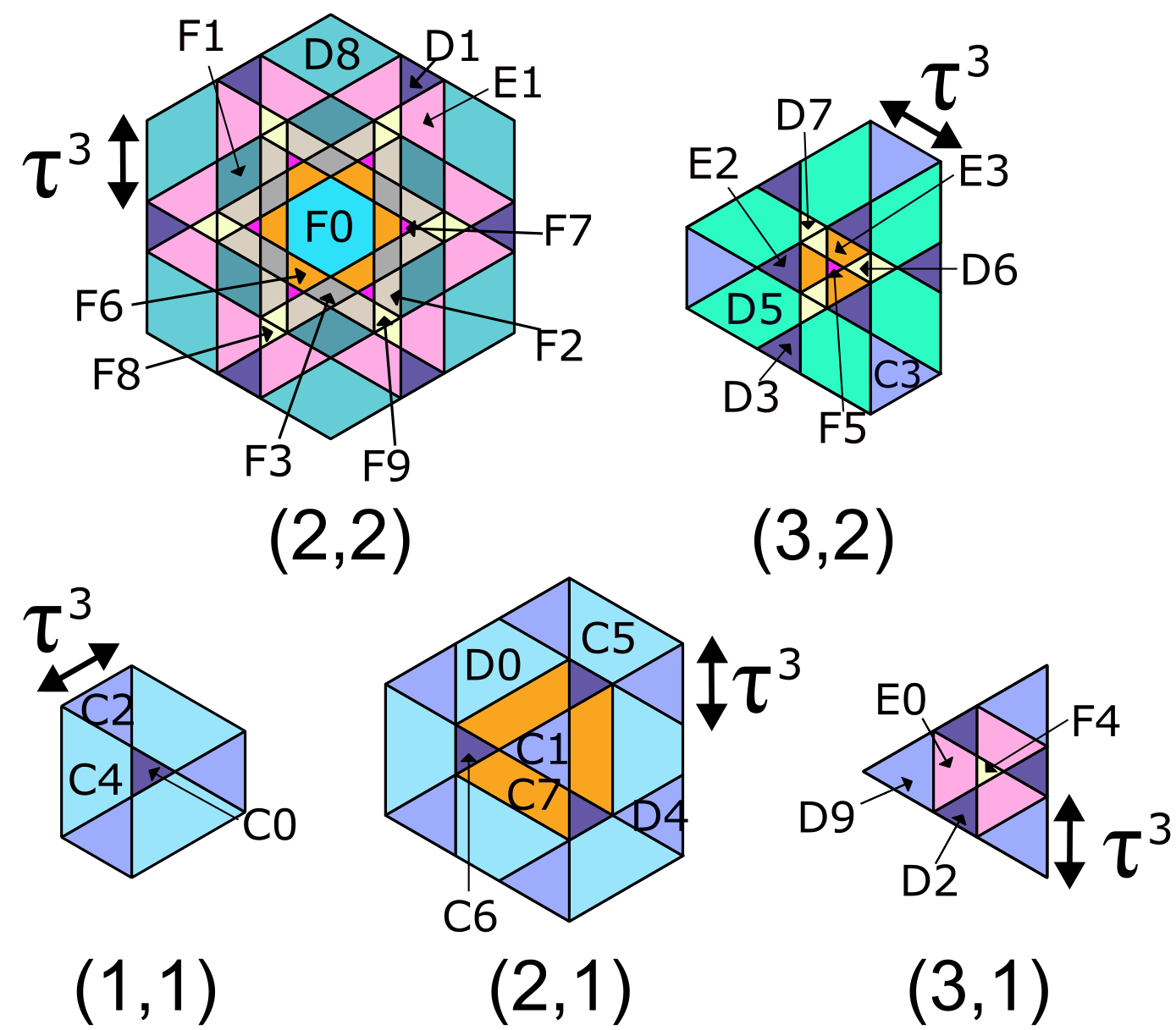}
	\caption{Five of the nine 2-dimensional cross sections through the internal-space window of the hexagonal $\Hhh$ tiling (as in Fig. \ref{fig:heights0505}), labeled by their level indices $(h_s,h_l)$, defined in Eqs.~\eqref{Eq:Height}. Not shown are the mirror reflections of the ones shown. Subdomains of the cross sections are labeled according to the vertex configurations, shown in Fig.~\ref{fig:0505_vertices}, that they generate, and color coded according to their areas. Double arrows labeled $\tau^3$ indicate the scale. Thus, if the smallest magenta triangles, $F5$ and $F7$, have unit area, then the yellow triangles, $D6$, $D7$, $F4$, $F8$, and $F9$, have area $\tau^2$, the dark purple triangles, $C0$, $C6$, $D1$, $D2$, $D3$, and $E2$, have area $\tau^4$, and the light purple triangles, $C1$, $C2$, $C3$, $D4$, and $D9$, have area $\tau^6$. All the remaining areas are simple sums and differences of these triangles. Note the structural similarity of the (1,1) and (3,1) cross sections with those of the $\Hzz$ windows of Fig.~\ref{fig:Perp00}.
 } \label{fig:Perp0505}
\end{figure}

\subsection{The hexagonal $\Hhh$ tiling}

\subsubsection{Basic properties}

The hexagonal $\Hhh$ tiling of Fig.~\ref{fig:Hexa55} consists of three prototiles as well---a large rhombus (LR), a small rhombus (SR), and a parallelogram (P)---where all the long edges are a factor of $\tau$ longer than the short ones. Again, the vertices of the hexagonal $\Hhh$ tiling come in two varieties---even or odd---inducing an imbalanced bipartitioning of the tiling. Here, the even vertices, colored black, originate from three different types of windows, with level indices (1,1) \& (3,3); (2,2); and (1,3) \& (3,1); while the odd vertices, colored white, originate from two types of windows, with level indices (2,1) \& (2,3); and (1,2) \& (3,2). The windows are shown in the inset of Fig.~\ref{fig:heights0505}, and reproduced with greater detail in Fig.~\ref{fig:Perp0505}. 

The frequencies of the two vertex types are therefore proportional to the total areas $A_{\text{even}} = (301\tau+186)$ and $A_{\text{odd}} = (359\tau+222)$, given in arbitrary
units, of the corresponding even-parity and odd-parity windows, respectively. These areas can directly be inferred from the internal-space windows in Fig.~\ref{fig:Perp0505} and their mirror reflections. Thus, the fractions of black and white vertices are given by 
\begin{subequations}\label{Eq:Frequency-EvenOdd}
\begin{align}
    f_\text{black} &= \frac{A_{\text{even}}}{A_{\text{even}}+A_{\text{odd}}}=\frac{1+2\sqrt{5}}{12}\simeq 0.4560, \label{Eq:feven}\\
    f_\text{white} &= \frac{A_{\text{odd}}}{A_{\text{even}}+A_{\text{odd}}}=\frac{11-2\sqrt{5}}{12}\simeq 0.5440.\label{Eq:fodd}
\end{align}
\end{subequations}

\subsubsection{Substitution rules}

\begin{figure*}
	\centering
	\includegraphics[width=\linewidth]{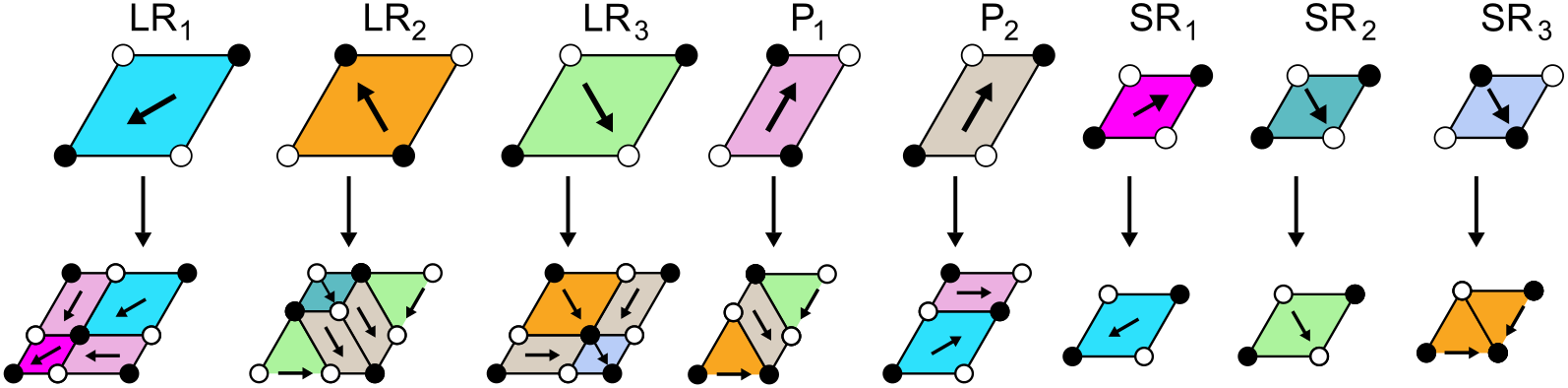}
	\caption{Substitution rules for the hexagonal $\Hhh$ tiling. To generate the tiling through substitution, the 3 tile scheme in Fig. \ref{fig:Hexa55} is expanded to 8 unique tiles, which are distinguished by colour. Top shows the initial tiles which are labelled with respect to their geometry (see text), and bottom indicates a generation of substitution. Black and white circles decorate vertices to highlight the bipartite sublattices which are discussed further in the text. Each tile (included those in the substitution) is marked with an arrow to indicate its relative orientation.}\label{fig:Deflation_0.5}
\end{figure*}

\begin{figure}
	\centering
	\includegraphics[width=\linewidth]{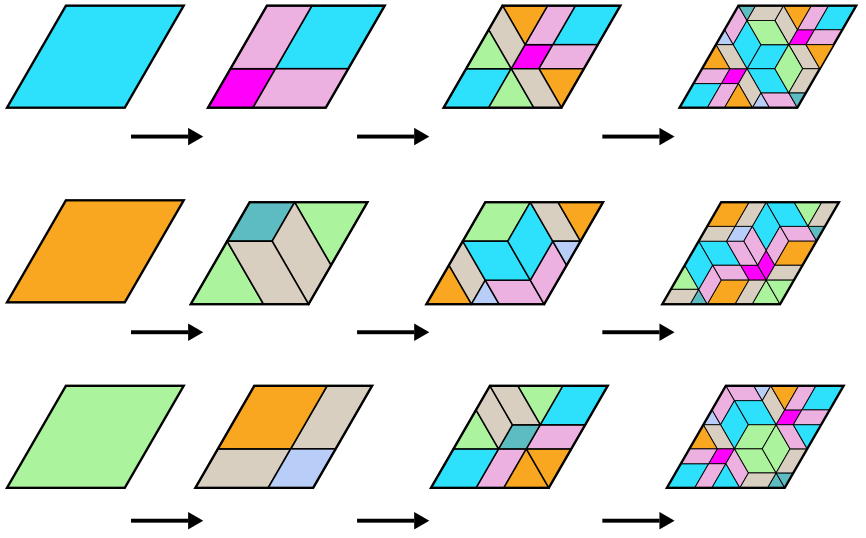}
	\caption{Three generations of substitution of the hexagonal $\Hhh$ tiling, starting from the three Large Rhombic tiles.} \label{fig:Deflation_0.5_examples}
\end{figure}

We are able to obtain substitution rules for the hexagonal $\Hhh$ tiling, through a process of trial-and-error, but only after expanding the set of prototiles from three to eight. These include three version of the large rhombus (LR$_1$, LR$_2$, LR$_3$), three versions of the small rhombus (SR$_1$, SR$_2$, SR$_3$), and two versions of the parallelogram (P$_1$, P$_2$), as shown on the top row of Fig. \ref{fig:Deflation_0.5}. Each of these eight tiles has a direction, indicated by arrows at their centers. These directions are required for specifying the substitution rules, as these break the 2-fold rotational symmetry of all the tiles. The substitution rules are shown on the bottom row of Fig.~\ref{fig:Deflation_0.5}. Three generations of substitution, starting with each of the three LR tiles, are shown in Fig.~\ref{fig:Deflation_0.5_examples}.

As with the hexagonal $H_{00}$ tiling, we can calculate the tile frequencies by means of the tile substitution-matrix,
\begin{equation} 
M_{\frac{1}{2}\frac{1}{2}} = 
  \kbordermatrix{
    & LR_1 & LR_2 & LR_3 & P_1 & P_2 & SR_1 & SR_2 & SR_3 \\
	LR_1 & 1  & 0   & 0   & 0   & 1  & 1  &  0  &  0  \\
	LR_2 & 0   & 0  &  1  & \frac{1}{2} &  0  & 0   & 0  &  1 \\[0.3em]
	LR_3 &  0  & 1   & 0  & \frac{1}{2}  &  0  &  0  &  1  & 0 \\[0.3em]
	P_1 & 2  &  0  &  0  & 0   & 1  &  0  &  0  &  0  \\
	P_2 &  0  & 2  & 2  & 1  &  0  &  0  &  0  &  0  \\
	SR_1 & 1  &  0  & 0   &  0  & 0   &  0  & 0   & 0   \\
	SR_2 & 0   & 1  & 0   & 0   &  0  &  0  &  0  & 0   \\
	SR_3 & 0   & 0   & 1  &  0  & 0   &  0  &  0  &  0 
  },
\end{equation}
whose largest eigenvalue is again $\tau^2$. The corresponding eigenvector is proportional to $(2\tau^2,\tau^2,\tau^2,4\tau,4\tau,,2,1,1)$.
Thus, there are an equal number of SR$_2$ and SR$_3$ tiles, yet twice as many SR$_1$ tiles, and so on. Considering all tiles of the same shape together, we find that the SR tile is the least frequent; LR appears $\tau^2=\tau+1$ more often; and the P tile, which is only slightly more frequent, appears $2\tau$ more often than the SR tile. These tile frequencies are consistent with an empirical inspection of the tiling in Fig.~\ref{fig:Hexa55}. Note that these are exactly the same frequencies one finds for the small-square, large-square, and rectangular tiles of the square Fibonacci tiling~\cite{squarefib}.

\begin{figure*}
	\centering
	\includegraphics[width=\linewidth]{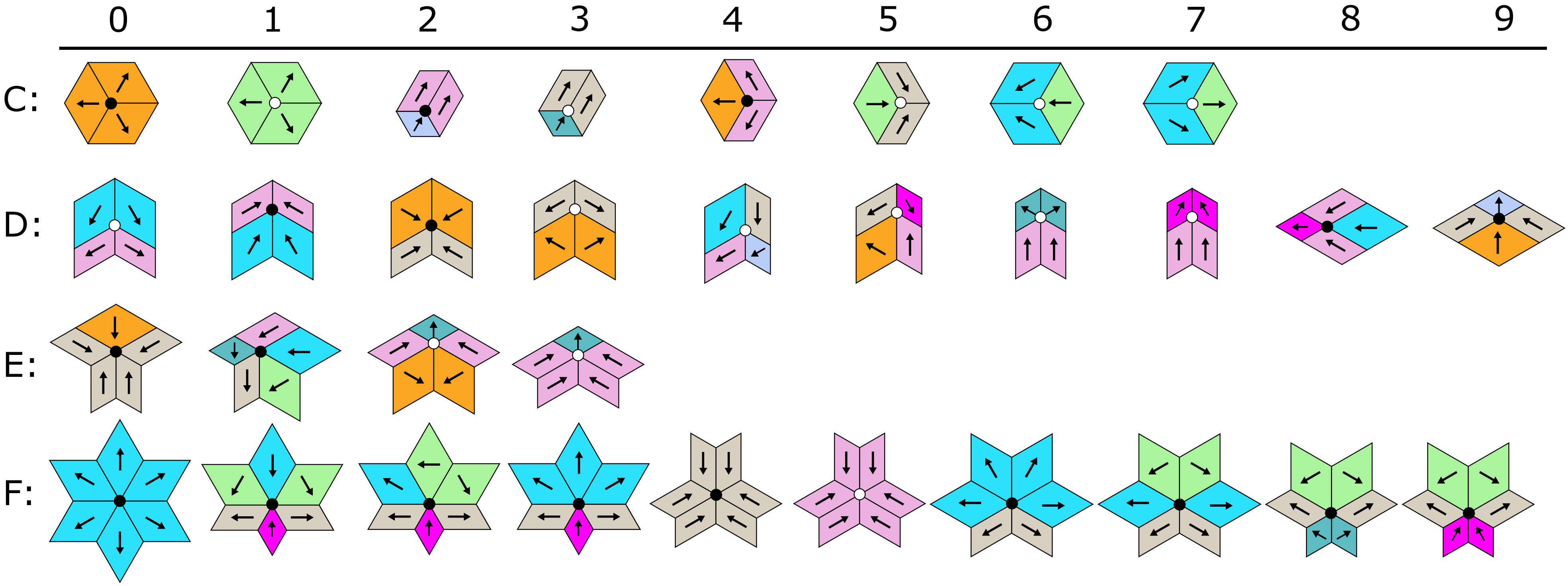}
	\caption{The 32 vertex configurations of the hexagonal $\Hhh$ tiling. The configurations are labeled by their coordination number---C through F---followed a numerical index. The vertices are colored---black or white---to indicate their parity. The tiles are decorated with arrows as in Fig.~\ref{fig:Deflation_0.5}.} 
    \label{fig:0505_vertices}
\end{figure*}

Considering the eight different tiles and their distinct orientations, we find 32 types of vertex configurations in the $\Hhh$ tiling, shown in Fig. \ref{fig:0505_vertices}. Each vertex configuration is labeled by a letter---from C to F---indicating its  coordination number---from 3 to 6, respectively. The letter is followed by a sequential numerical index.
The subdomains that these vertices occupy within the internal-space windows are labelled accordingly in Fig.~\ref{fig:Perp0505}, and can be used directly to calculate the frequencies of the 32 different vertex configurations in the tiling, summarized in Table~\ref{table}. Again, the same frequencies are obtained by constructing the vertex substitution-matrix $V_{\frac12\frac12}$ for the tiling and finding the eigenvector that corresponds to the largest eigenvalue. These calculations, which are too lengthy to describe here, lead to the same distribution of vertex configurations, given in Table~\ref{table}, 


\begin{table}
\caption{Frequencies of the 32 vertex configurations of the hexagonal $\Hhh$ tiling, shown in Fig.~\ref{fig:0505_vertices}.}
    \begin{tabular}{c|cccccccccc}
    \hline
    \hline
    &0&1&2&3&4&5&6&7&8&9\\
    \hline
    C&$\displaystyle\frac{1}{12\tau^6}$&$\displaystyle\frac{1}{12\tau^4}$&
    $\displaystyle\frac{1}{4\tau^4}$&$\displaystyle\frac{1}{4\tau^4}$&
    $\displaystyle\frac{\sqrt{5}}{4\tau^4}$&$\displaystyle\frac{\sqrt{5}}{4\tau^4}$&
    $\displaystyle\frac{1}{4\tau^6}$&$\displaystyle\frac{1}{4\tau^3}$&\\[1em]
    D&$\displaystyle\frac{\sqrt{5}}{4\tau^4}$&$\displaystyle\frac{1}{4\tau^6}$&
    $\displaystyle\frac{1}{4\tau^6}$&$\displaystyle\frac{1}{4\tau^6}$&
    $\displaystyle\frac{1}{2\tau^4}$&$\displaystyle\frac{3}{2\tau^5}$&
    $\displaystyle\frac{1}{4\tau^8}$&$\displaystyle\frac{1}{4\tau^8}$&
    $\displaystyle\frac{1}{2\tau^4}$&$\displaystyle\frac{1}{4\tau^4}$\\[1em]
    E&$\displaystyle\frac{\sqrt{5}}{4\tau^6}$&$\displaystyle\frac{\sqrt{5}}{2\tau^6}$&
    $\displaystyle\frac{1}{4\tau^6}$&$\displaystyle\frac{\sqrt{5}}{4\tau^8}$&\\[1em]
    F&$\displaystyle\frac{1}{4\tau^6}$&$\displaystyle\frac{1}{2\tau^6}$&
    $\displaystyle\frac{\sqrt{5}}{2\tau^8}$&$\displaystyle\frac{1}{2\tau^8}$&
    $\displaystyle\frac{1}{12\tau^8}$&$\displaystyle\frac{1}{12\tau^{10}}$&
    $\displaystyle\frac{\sqrt{5}}{4\tau^8}$&$\displaystyle\frac{1}{4\tau^{10}}$&
    $\displaystyle\frac{1}{4\tau^8}$&$\displaystyle\frac{1}{4\tau^8}$\\[.8em]
    \hline
    \hline
    \end{tabular}
\label{table}
\end{table}



\section{Summary and outlook}
\label{sec:summary}

Driven by the fruitful use of nonminimal-rank tilings with tetragonal symmetry~\cite{squarefib} as theoretical models for studying the physics of quasicrystals~\cite{Shahar06, *Shahar08, Damanik11}, and motivated by the experimental observation of aperiodic crystals with trigonal and hexagonal symmetry~\cite{Coates20,Woods14,Uri23}, we have introduced a two-parameter family of 2-dimensional rank-4 trigonal tilings, four members of which are hexagonal. We have constructed and analyzed this family of tilings using both the cut-and-project and the dual-grid methods, and have provided substitution rules for generating two of its hexagonal members.

The tilings studied here are part of a broader family of 2-dimensional rank-4 trigonal and hexagonal tilings, generated by a pair of mutually incommensurate 6-fold stars of wave vectors in reciprocal space. Here we have focused on the special case where the two stars are co-aligned ($\theta=0$), and where the length ratio $\tau$ of vectors from the two stars is the golden mean $(1+\sqrt{5})/2$. This ratio produces particularly aesthetic tilings, and is suitable for studying the 3-fold surfaces of icosahedral quasicrystals. 

\begin{figure*}
    \centering
    \begin{subfigure}[b]{0.48\textwidth}
         \centering
         \includegraphics[width=\textwidth]{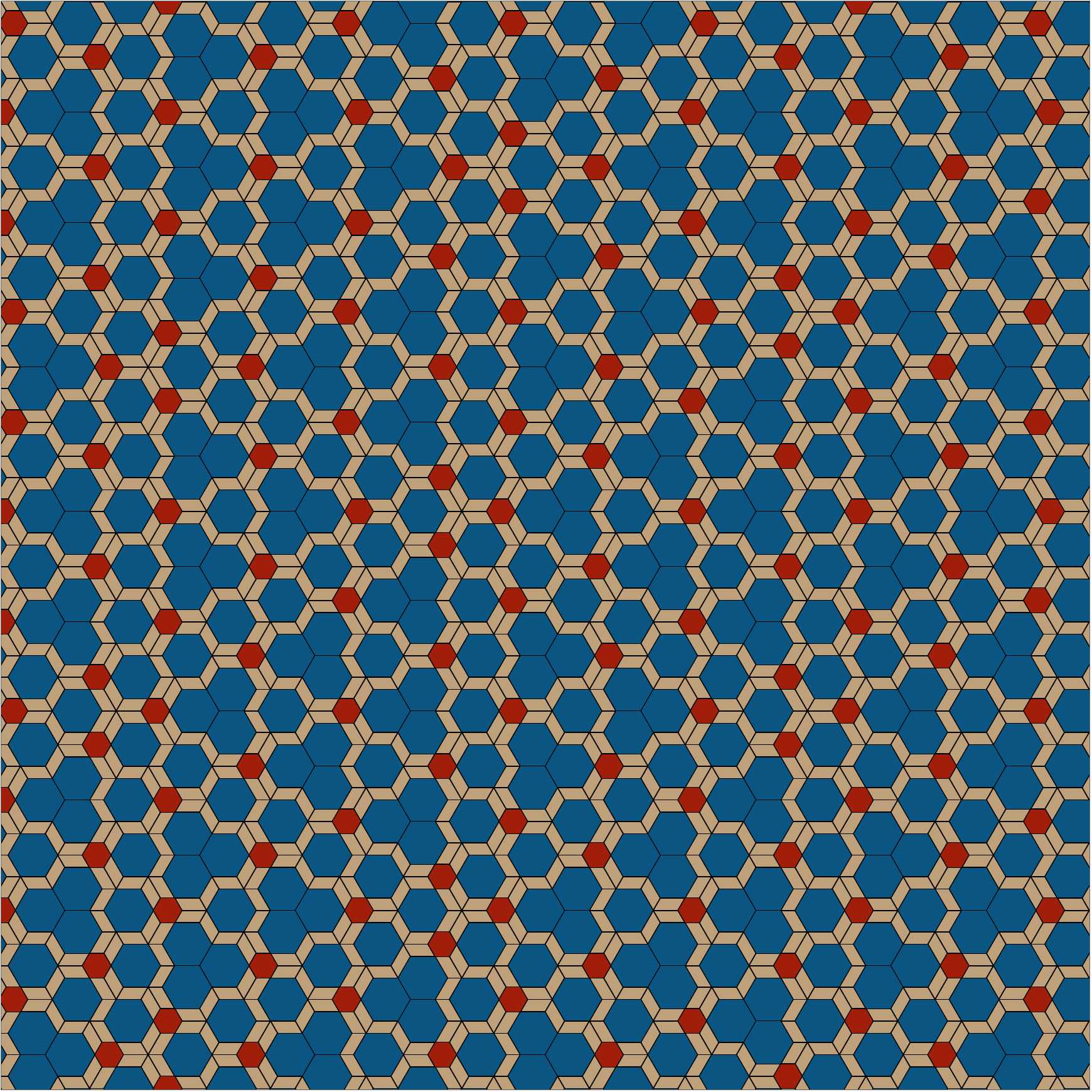}
         \caption{Rank-4 hexagonal tiling with $\tau=\sqrt{3}$ and $\alpha_s\equiv\alpha_l\equiv0$}
         \label{fig:HexaRoot3}
     \end{subfigure}
     \hfill
     \begin{subfigure}[b]{0.48\textwidth}
         \centering
         \includegraphics[width=\textwidth]{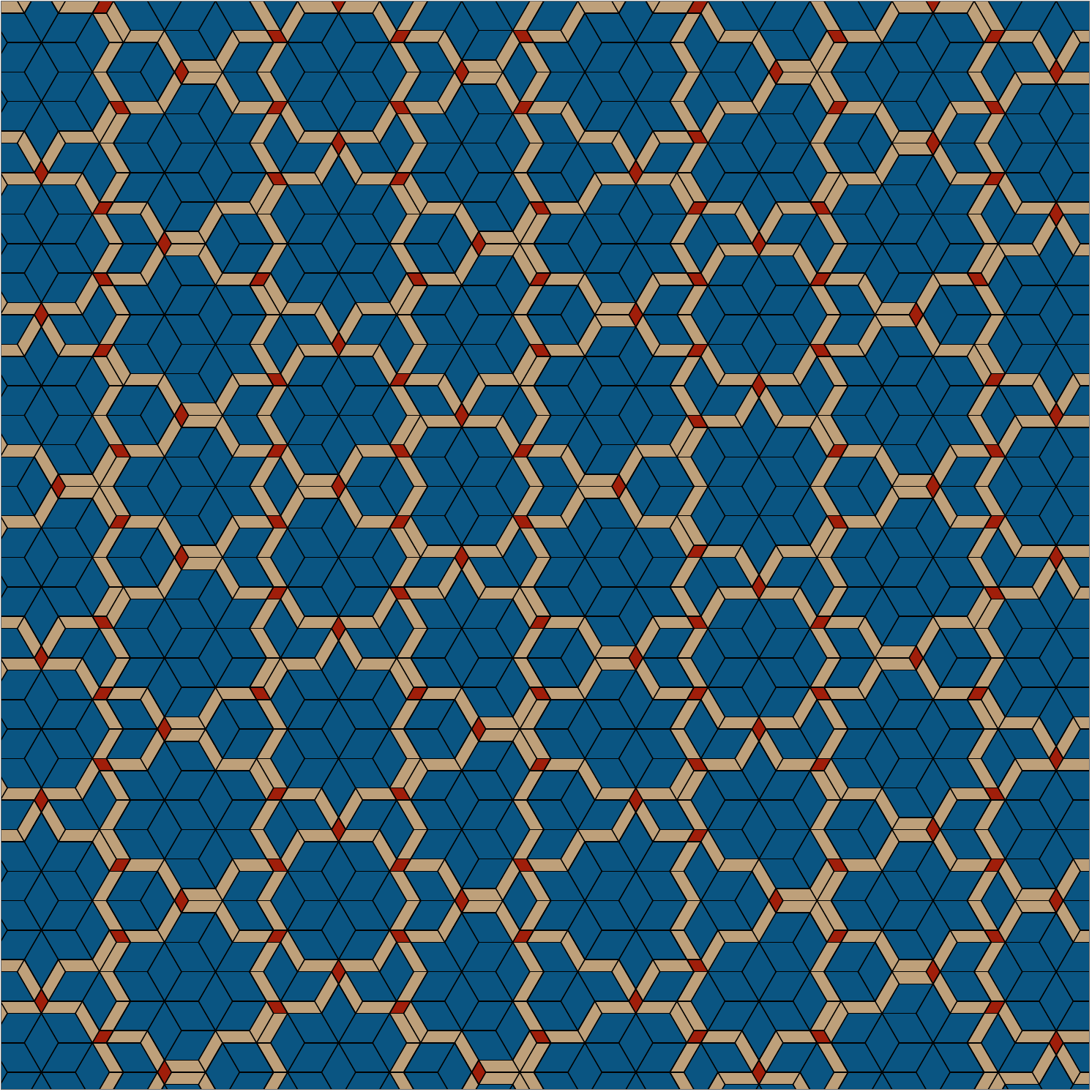}
         \caption{Hexagonal silver-mean tiling with $\alpha_s\equiv\alpha_l\equiv0.5$}
         \label{fig:HexaSilver55}
     \end{subfigure}
    \begin{subfigure}[b]{0.48\textwidth}
         \centering
         \includegraphics[width=\textwidth]{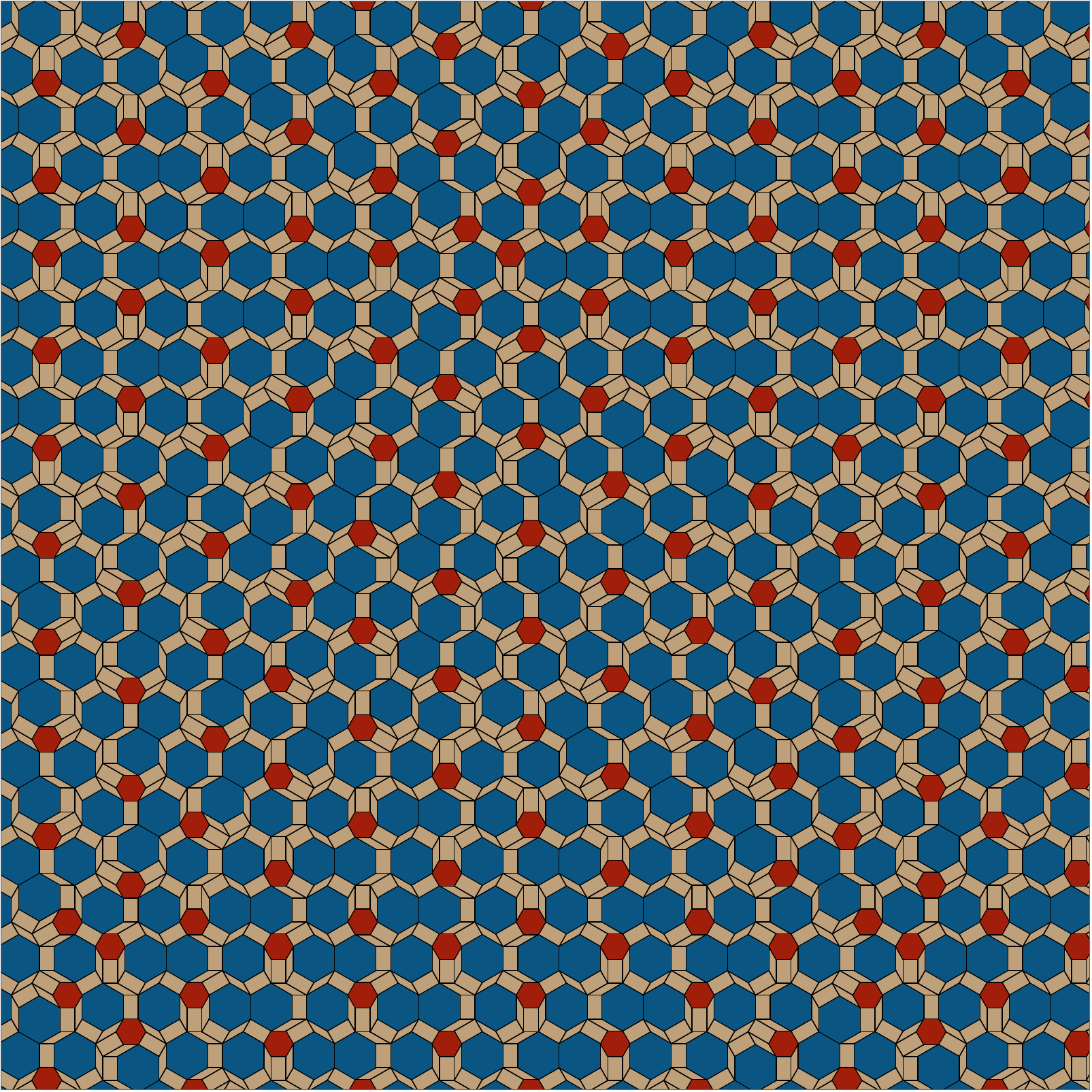}
         \caption{Hexagonal golden-mean tiling with $\alpha_s\equiv\alpha_l\equiv0$, and $\theta=30^\circ$}
         \label{fig:HexaRotatedGolden00}
     \end{subfigure}
     \hfill
     \begin{subfigure}[b]{0.48\textwidth}
         \centering
         \includegraphics[width=\textwidth]{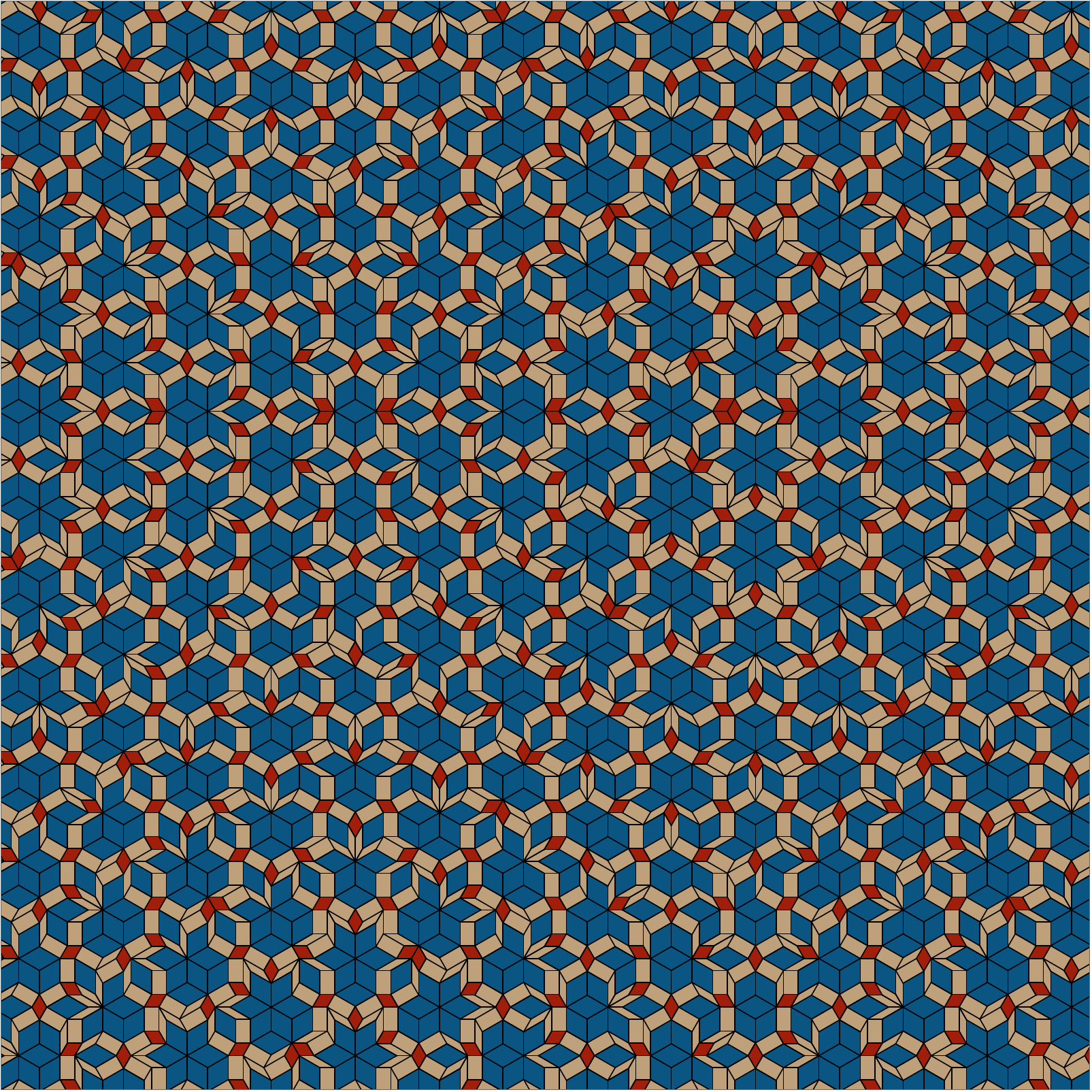}
         \caption{Hexagonal golden-mean tiling with $\alpha_s\equiv\alpha_l\equiv0.5$, and $\theta=30^\circ$}
         \label{fig:HexaRotatedGolden55}
     \end{subfigure}
    \caption{Top row: Rank-4 hexagonal tilings, obtained by the dual-grid method, as in Fig.~\ref{Fig:Hexa-tilings}, but with $\tau$ set to the values of $\sqrt{3}$, and the silver mean $1+\sqrt{2}$ (a metallic-mean case which has been discussed further in \cite{Matsubara23}). The case of $\tau=\sqrt{3}$ is intriguing because the diffraction diagram (not shown here) contains 12-fold symmetric rings of Bragg peaks, whose intensities alternate, thus exhibiting only 6-fold symmetry.
    Bottom row: Hexagonal Fibonacci, or golden-mean tilings, with $\tau$ still set to the value of the golden mean $(1+\sqrt{5})/2$, but with the two trigrids rotated by $30^\circ$ with respect to each other. One can see that the large and small hexagonal or rhombic tiles are now rotated with respect to each other. Accordingly, in order to fill in the gaps between these tiles, there are now two parallelogram tiles---a $30^\circ$ parallelogram and a rectangle---instead of the single $60^\circ$ parallelogram. }
    \label{Fig:Generalized-tilings}
\end{figure*}

Generalizations are numerous, and merely require one to change the values of the parameters $\tau$ and $\theta$, and repeat the procedures of sections~\ref{sec:dual} and~\ref{sec:projection}, above. Any irrational $\tau$ like $\pi$ or $e$ will do, but certain values are particularly interesting. For example, setting $\tau=\sqrt{3}$, shown in Fig.~\ref{fig:HexaRoot3}, produces a tiling whose diffraction diagram (not shown here) contains 12-fold symmetric rings of Bragg peaks, whose intensities alternate, thus exhibiting only 6-fold symmetry. One can also directly obtain the full sequence of rank-4 hexagonal or trigonal metallic-mean tilings, simply by setting the ratio $\tau$ to $(n+\sqrt{n^2+4})/2$ ($n\in\mathbb{N}$) in Eqs.~\eqref{Eq:Lengths} and \eqref{Eq:a-star}. For example, Fig.~\ref{fig:HexaSilver55} shows the hexagonal silver-mean tiling with $\alpha_s\equiv\alpha_l\equiv0.5$. The generation of hexagonal metallic-mean tilings with $\alpha_s\equiv\alpha_l\equiv0$ (not shown here), using substitution rules, has recently been treated elsewhere by two of us~\cite{Matsubara23}. Alternatively, one could substitute experimentally measured lattice parameters and relative rotation angles $\theta$ for studying real structures like graphene on hexagonal boron nitride. Two additional generalizations, obtained by keeping $\tau$ as the golden mean, but applying a nonzero rotation $\theta=30^\circ$, are shown in Figs.~\ref{fig:HexaRotatedGolden00} and \ref{fig:HexaRotatedGolden55}. Possible further generalizations might involve the increase of the rank by introducing a third trigrid, or perhaps even more, if that would ever become interesting experimentally.  

\begin{acknowledgments}
This work was supported by Grant-in-Aid for Scientific Research from JSPS,
KAKENHI Grants
No. JP17K05536, JP19H05821, JP21H01025, JP22K03525 (A.K.), and
No. JP19H05817 and No. JP19H05818 (S.C.); the EPSRC grant EP/X011984/1 (S.C); and by the Israel Science Foundation (ISF) through grant No.~1259/22 (R.L.). S.C.\ and R.L.\ would like to thank the late Prof.\ Uwe Grimm for his inspiration and encouragement in the development of trigonal and hexagonal aperiodic tilings. R.L.\ also thanks Ronan McGrath for his kind hospitality during his extended stay at the Dept.\ of Physics at the University of Liverpool as a Leverhulme Visiting Professor. 
\end{acknowledgments}

\bibliography{HexaBib}
\end{document}